\documentclass{aa}  

\usepackage{txfonts}
\usepackage{tabularx}
\usepackage{subfigure}  
\usepackage{graphicx,amsmath}
\usepackage{amssymb}
\usepackage{xcolor}
\usepackage{natbib}      
\usepackage{url}
\usepackage{multirow}
\usepackage[para,online,flushleft]{threeparttable}   
\usepackage{soul,mathabx}
\bibpunct{(}{)}{;}{a}{}{,} % to f
\usepackage{lscape}
\usepackage{adjustbox,textcomp}
\usepackage{dsfont}
\usepackage{stfloats}

\usepackage{hyperref}   
\hypersetup{
    colorlinks=true,
    linkcolor=blue,
    filecolor=blue,      
    urlcolor=blue,
    citecolor=blue,
    }
\usepackage{braket}     
\usepackage{upgreek}    
\usepackage{array,multirow,booktabs}

\newcommand{\J}{\texttt{JADE}}

\begin{document}

   \title{The JADE\thanks{This work has been performed using the following software: \url{https://github.com/JADE-Exoplanets/JADE}. Another distribution exists at: \url{https://gitlab.unige.ch/spice_dune/jade}.} code. II. Modeling the coupled orbital and atmospheric evolution of GJ~436~b to constrain its migration and companion}

   \author{M.~Attia\inst{1}
      \and V.~Bourrier\inst{1}
      \and E.~Bolmont\inst{1}
      \and L.~Mignon\inst{2}
      \and J.-B.~Delisle\inst{1}
      \and H.~Beust\inst{2}
      \and N.~C.~Hara\inst{3}
      \and C.~Mordasini\inst{4,5}
          }

   \institute{Observatoire Astronomique de l'Universit\'e de Gen\`eve, Chemin Pegasi 51b, 1290 Versoix, Switzerland\\
              \email{mara.attia@unige.ch} 
         \and Univ. Grenoble Alpes, CNRS, IPAG, 38000 Grenoble, France
         \and Aix Marseille Universit\'e, CNRS, CNES, LAM, Marseille, France
         \and Space Research and Planetary Sciences, Physics Institute, University of Bern, Gesellschaftsstrasse 6, 3012 Bern, Switzerland
         \and Center for Space and Habitability, University of Bern, Gesellschaftsstrasse 6, 3012 Bern, Switzerland
             }

\authorrunning{M.~Attia et al.}
\titlerunning{The JADE code: Modeling and constraining the history of GJ~436}

\date{Received 21 April 2025 / Accepted 25 August 2025}

\abstract
   {
   
The observed architecture and modeled evolution of close-in exoplanets provide crucial insights into their formation pathways and survival mechanisms. To investigate these fundamental questions, we employed Joining Atmosphere and Dynamics for Exoplanets (\J{}), a comprehensive numerical code that self-consistently models the coupled evolution of planetary atmospheres and orbital dynamics over secular timescales, rooted in present-day observations. \J{} integrates atmospheric photoevaporation with high-eccentricity migration processes driven by von Zeipel--Lidov--Kozai (ZLK) cycles from an external perturber, allowing us to explore evolutionary scenarios where dynamical and atmospheric processes influence each other. Here, we specifically considered GJ~436~b, a warm Neptune with an eccentric orbit and polar spin--orbit angle that has survived within the ``hot Neptune desert'' despite ongoing atmospheric escape. Our extensive exploration of parameter space included over $500\,000$ fully coupled \J{} simulations in a framework that combines precomputed grids with Bayesian inference. This allowed us to constrain GJ~436~b's initial conditions and the properties of its putative perturbing companion within a ZLK hypothesis. Our results suggest that GJ~436~b formed at $\sim\,0.3\,\mathrm{AU}$ and, despite its current substantial atmospheric erosion, has experienced minimal cumulative mass loss throughout its history, thanks to a late inward migration triggered by a distant companion inducing ZLK oscillations. We find that initial mutual inclinations of $80^\circ - 100^\circ$ with this companion best reproduce the observed polar orbit. By combining our explored constraints with radial velocity detection limits, we identified the viable parameter space for the hypothetical GJ~436~c. We found that it strongly disfavors stellar and brown dwarf masses, which offers a useful guide for future observational searches. This work demonstrates how coupled orbital--atmospheric modeling can shed light on the complex interplay of processes shaping close-in exoplanets and explain the survival of volatile-rich worlds near the edges of the hot Neptune desert.

   }


\maketitle

\section{Introduction}
\label{sec:intro}

Exoplanet research has rapidly expanded our understanding of planetary systems beyond our own \citep{Winn2015}. Astronomers have discovered over $5\,800$ exoplanets so far, with nearly half occupying extremely close orbits around their host stars, completing their orbital revolutions in less than ten days (NASA Exoplanet Archive, consulted in April 2025). This population of close-in planets demonstrates remarkable diversity, ranging from small rocky worlds to giant gas planets. However, a notable and intriguing feature emerges within this population: a scarcity of Neptune-sized planets with orbital periods shorter than approximately three days. This phenomenon is known as the ``hot Neptune desert'' \citep[e.g.,][]{Lecavelier2007,Szabo2011,Beauge2012,Lundkvist2016}.

Despite extensive research over the past decade, the fundamental origins and precise characteristics defining the boundaries of this desert are not entirely understood. Two primary hypotheses have emerged to explain this planetary population distribution. First, atmospheric evaporation driven by intense stellar radiation could play a crucial role \citep[e.g.,][]{Owen2018}, potentially transforming Neptune-like planets into smaller, rocky bodies \citep{Lecavelier2004,Owen2013,Lopez2013}. Stellar X-ray and extreme ultraviolet (XUV) radiation can induce hydrodynamic atmospheric escape, particularly affecting planets in close proximity to their host stars \citep[e.g.,][]{Lammer2003,MurrayClay2009,Tripathi2015,Owen2019,Vissapragada2022}. Second, alternative theories propose that orbital migration mechanisms, including disk-driven migration \citep[e.g.,][]{Goldreich1979,Lin1996,Baruteau2016}, secular chaos \citep[e.g.,][]{Wu2011,Hamers2017}, planet--planet scattering \citep[e.g.,][]{Ford2008,Nagasawa2008}, and von Zeipel--Lidov--Kozai \citep[ZLK,][]{vonZeipel1910,Lidov1962,Kozai1962} migration, might significantly influence planetary system architectures and composition.

To better understand the interaction between orbital migration and atmospheric escape for close-in exoplanets, there is a need for dedicated models to trace these processes over time, grounded in today's observational constraints. While $N$-body integrators are highly accurate in simulating dynamical evolution \citep[e.g.,][]{Rein2012,Bolmont2020,Trani2023}, they often neglect planetary structure and fail to monitor long-term changes, limiting their ability to fully capture the history of older systems. On the other hand, atmospheric models provide detailed snapshots of planetary envelopes \citep[e.g.,][]{Heng2012,Komacek2017,Johnstone2018} but struggle to account for temporal evolution and migration. The Joining Atmosphere and Dynamics for Exoplanets (\J{}) code \citep{Attia2021} was developed to address these limitations by integrating both orbital and atmospheric evolution, particularly focusing on late-stage migration scenarios in hot and warm Neptunes influenced by ZLK cycles. \J{} is designed to simulate the intricate interplay between atmospheric erosion, orbital changes, and planetary structure in real time, offering a more comprehensive understanding of these phenomena in giant exoplanets. 

In the first article \citep{Attia2021}, we benchmarked the \J{} code on the intriguing case of GJ~436~b \citep{Butler2004}, a Neptune-sized planet still located at the fringes of the desert despite its current evaporation at tremendous rates \citep{Bourrier2016}. This planet further stands out because of its eccentric orbit \citep{Trifonov2018}, which should have been circularized a long time ago by tidal forces raised by the close star (as later discussed in Sect.~\ref{sect:motiv}). A possible scenario adduces a hidden distant companion generating a ZLK resonance resulting in a late migration \citep{Bourrier2018,Attia2021}. By forming farther out than its current position, GJ~436~b would have been trapped in the resonance for billions of years and migrated to a close-in orbit only recently; hence, it would have escaped the intense irradiation of the star when it was young and energetic \citep[e.g.,][]{Jackson2012}, thereby explaining its survival inside the desert. This narrative also naturally explains the residual eccentricity, as well as its polar spin--orbit angle \citep{Bourrier2022}, given the pervasive tendency of high-eccentricity migration mechanisms (HEM, which encompass ZLK) to elongate orbits and excite inclinations \citep[e.g.,][]{Fabrycky2009,Nagasawa2011}. 

In the process, we revealed, for the first time in the context of ZLK cycles, a strong mutual feedback between the orbital and atmospheric history. Specifically, during the resonance, near-unity eccentricities periodically bring the planet in proximity to the star, heating and inflating its atmosphere to a significant extent, thereby inducing radius pulsations that are in tune with the orbital oscillations. While previous studies have explored coupled dynamics--atmosphere evolution involving radius inflation \citep[e.g.,][]{Gu2003,Miller2009,Lopez2016,Glanz2022}, the specific phenomenon of radius pulsations driven by the extreme eccentricity variations during ZLK cycles arising from the secularization of irradiation power that increases dramatically as $e \rightarrow 1$ \citep{Attia2021} has not been previously investigated. This is essential to account for in any secular investigation because of the substantial long-term effects of a varying radius in controlling the atmospheric structure as well as the intensity of tidal effects.

In light of these promising results, in this second article, we have resumed our the investigation of GJ~436~b, whose history might be prototypical of the evolution of intermediate-size planets at the periphery of the desert. Since the seminal definition of its borders by \citet{Mazeh2016}, its inner edges have become less and less desertic, which led \citet{CastroGonzalez2024} to update them, unveiling a surprising overdensity of warm Neptunes at the desert boundary (the so-called Neptunian ``ridge''). As in the case of GJ~436~b, it is noteworthy that most planets populating this ridge possess eccentric orbits \citep{Correia2020}, coupled with substantial spin--orbit tilts \citep{Bourrier2023,Attia2023}. If all exo-Neptunes migrated early within their protoplanetary disk, we would expect the ridge to be dominated by circular and aligned systems. This is because tidal circularization and realignment are more effective at shorter periods, extinguishing any moderate primordial ellipticity and misalignment caused by disk-driven migration \citep[e.g.,][]{Marzari2009}. The ridge could consequently be the stable endpoint of the tumultuous HEM transporting Neptunian worlds from their supposed birthplace beyond the ice line, thereby solving the puzzle of the survival of their eroding atmospheres. 

Assessing this scenario requires deploying advanced fully joint models of dynamical--atmospheric evolution for a large selection of Neptune-sized planets, which is made possible by the \J{} code. While the first article \citep{Attia2021} demonstrated the feasibility of the HEM-protracted erosion scenario for GJ~436~b, this study goes beyond by quantitatively constraining the system's evolutionary pathway through a comprehensive parameter space exploration. Specifically, we seek to determine: (1) the planet's formation location and initial mass; (2) the properties of the hypothetical companion GJ~436~c, providing observationally testable predictions for its mass and orbital period; and (3) the range of initial mutual inclinations compatible with the observed polar orbit. If a solution that is compatible with today's observations would emerge, our narrative would be validated as viable, perhaps even hinting at similar histories for other ridge tenants. In Sect.~\ref{sect:jade}, we recapitulate the building blocks of the \J{} code and present its new features. Section~\ref{sect:preproc} explains the preprocessing steps facilitating the application of \J{} to GJ~436~b for its parameter space exploration, with the latter  described in Sect.~\ref{sect:explo}. Our results are then used to pinpoint the possible location of GJ~436~c within our ZLK framework in Sect.~\ref{sect:detect}, while accounting for the existing observational constraints. We present our conclusions in Sect.~\ref{sect:conclu}.

\section{The \J{} code}
\label{sect:jade}

\subsection{Overview}

The \J{} code \citep{Attia2021} simulates the evolution of a planet over secular timescales of several billion years (Gyrs) under the combined influence of realistic dynamical and atmospheric mechanisms. The dynamical evolution is governed by the action of an outer, distant third body in the system, capable of driving a ZLK orbital resonance of the close-in planet \citep[e.g.,][]{Naoz2016}, counterbalanced by short-range forces (SRFs, encompassing tidal effects, spin distortion, and general relativity) binding the inner binary and quenching the resonance \citep[e.g.,][]{Liu2015}. The planet is modeled as composed of an iron core, a silicate mantle, and a H/He gaseous envelope, whose varying properties are self-consistently derived at each time step as the orbit, stellar irradiation, and inner planetary heating all evolve. The atmosphere easily erodes due to XUV-induced photoevaporation \citep[e.g.,][]{Owen2019} and internal released luminosity \citep[e.g.,][]{Ginzburg2016}, and \J{} calculates this mass-loss rate using analytical formulae adjusted to detailed simulations of upper atmospheric structures \citep{Salz2016}. This provides a better agreement with observed mass-loss rates than the commonly used energy-limited approximation \citep[e.g.,][]{Watson1981,Lammer2003,Erkaev2007}. Appendix~\ref{app:jade_101} illustrates the overall functioning of the \J{} code.

\subsection{Updates and new features}
\label{sect:jade_new}

Since the initial release of the \J{} code \citep{Attia2021}, we have implemented several new features for a more comprehensive and realistic treatment of close-in gas giants. While some of these improvements have been utilized in intermediate studies, we present here for the first time a complete description of all major updates, establishing this work as the authoritative reference for the current version of \J{}. Accounting for the stellar Roche limit was the first of them, and succeeding \J{} analyses insisted on its importance especially given the extreme adjacency of our targets to their hosts \citep[e.g.,][]{Pezzotti2021,Almenara2022}. Formally speaking, any simulation halts if the planet overflows inside the stellar Roche limit; in other words, if the following condition is reached

\begin{equation}
\label{eq:roche}
a\left(1 - e\right) < 2 \times R_\mathrm{p} \left(\frac{M_\star}{M_\mathrm{p}}\right)^{1/3}.
\end{equation}

\noindent The left-hand side corresponds to the distance at periastron, enforced by the $\left(1 - e\right)$ factor, as tidal forces are maximal at that point of the orbit. The factor of 2 on the right-hand side is a conservative value to account for the fluidity of the tidally disruptable material \citep[e.g.,][]{Chandrasekhar1963,Teyssandier2019}. 

Moreover, a more rigorous calculation of the outer boundary pressure (i.e., the pressure at the photosphere or transit radius, where the integration starts) replaces the arbitrary former value of $P_\mathrm{tr} = 1\,\mathrm{mbar}$. The procedure is the following: choosing a certain $P_\mathrm{tr}$ will set the gas density at the photosphere via the equations of state (EoS), which, in turn, will set an outer boundary opacity, $\kappa_\mathrm{tr}$, through the collection of tabulated opacities \J{} employs \citep{Ferguson2005}. Plus, the radiative conditions at the transit radius justify that $P_\mathrm{tr}$ is none other than the photospheric pressure\footnote{In other terms, we neglect the radiation pressure emanating from the planet's interior.}

\begin{equation}
\label{eq:Ptr}
P_\mathrm{tr} = \frac{\tau g}{\kappa_\mathrm{tr}},
\end{equation}

\noindent where $g$ is the surface gravity of the planet and $\tau$ is the optical depth (equal to 2/3 at the photosphere, following the Eddington approximation). Hence, $P_\mathrm{tr}$ is hence iterated over until convergence of Eq.~(\ref{eq:Ptr}), which is carried out every time the structure integrator is called since the outer boundary temperature dynamically evolves.

The layered prescription for the planet structure has also been improved. While the atmosphere remains unchanged, divided into an upper region collecting the high-energy incident flux and a lower region redistributing energy via radiation and convection, the solid material is now more faithful to the reality of planet interiors. It can be broken down into an iron core ($\alpha$-Fe, ferrite) topped by a silicate mantle (MgSiO$_3$, perovskite), which is well-described by a temperature-independent modified polytropic EoS \citep{Seager2007}. Such a composition is reminiscent of Earth and provides a good trade-off between a simple--fast model and an honest agreement with complex interior models \citep[e.g.,][]{Dorn2021}.

\begin{figure*}
\includegraphics[width=\linewidth]{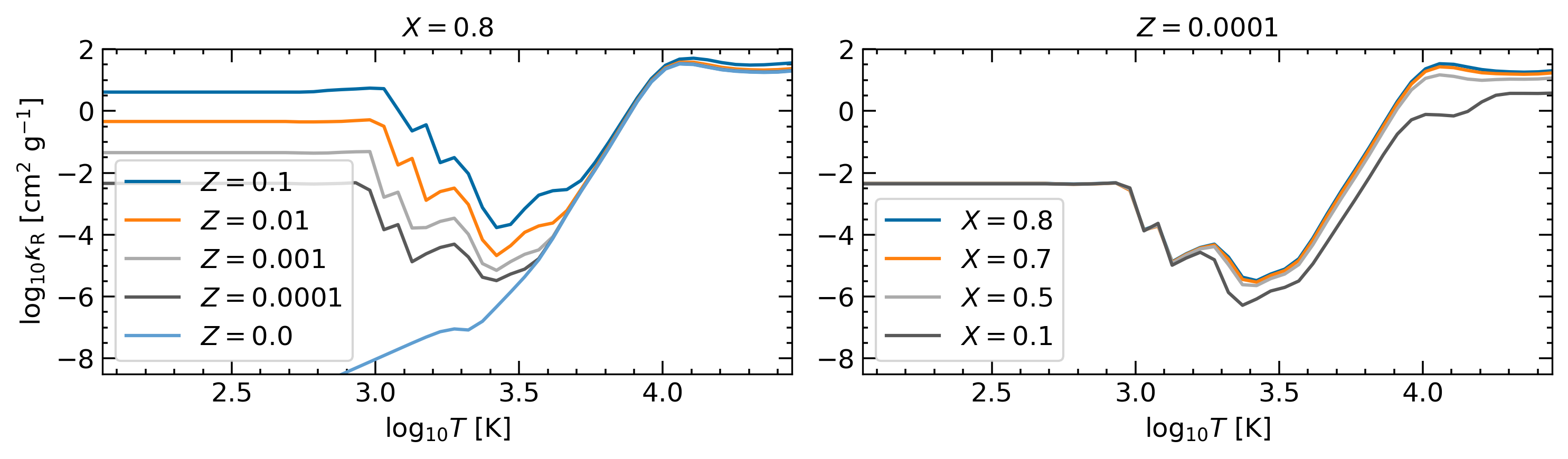}
\centering
\caption{Rosseland mean opacity, $\kappa_\mathrm{R}$, as a function of the temperature, $T$, interpolated from the tables of \citet{Ferguson2005}.
\textbf{Left:} Varying the metallicity, $Z$, at a fixed hydrogen mass ratio $X = 0.8$.
\textbf{Right:} Varying the hydrogen mass ratio, $X$, at a fixed metallicity $Z = 0.0001$.}
\label{fig:opacity_z}
\end{figure*}

The final major update concerns the opacity treatment in the atmosphere. The \J{} code employs the preferred formalism for such one-dimensional (1D) models: the Rosseland mean opacity \citep{Rosseland1924}, which represents a weighted average across many monochromatic opacities. It simplifies the process by enabling models to perform a single calculation for all frequencies instead of adjusting the calculation for each one. Considering the complex impact of a gas on a spectrum, this approach makes the models far more tractable. Atmospheres simulated by the \J{} code are composed of H/He, and the novelty now is that a trace amount of metals can also be introduced in the gaseous material, which leads to a massive change in the associated opacities. This is best visible in Fig.~\ref{fig:opacity_z}, where the Rosseland mean opacity $\kappa_\mathrm{R}$ is represented as a function of the gas temperature $T$ for a variety of compositions, as interpolated from the tabular entries used by \J{} \citep{Ferguson2005}. The left panel shows the influence of the atmospheric metal mass fraction $Z$, especially explicit in the planetary regime (lowest temperatures). Adding even a hint of heavy elements pumps the opacity from a negligible value to a physical one. If the presence of metals does not seem to be important for the hottest atmospheres, low temperatures on the other hand inhibit collision-induced absorption, making heavy dust grains absorbing light the main source of opacity. This is corroborated by the linear increase in $\kappa_\mathrm{R}$ with $Z$, seen for $T < 1\,000\,\mathrm{K}$. On the contrary, the right panel tells us the minimal sensitivity of $\kappa_\mathrm{R}$ to the hydrogen fraction $X$, all other things being equal, except for the highest temperatures, even though marginally. Indeed, atomic absorption becomes the biggest contributor to the opacity for the hottest gases, while a lower $X$ means more helium $Y$, which translates into helium lines quickly dominating over hydrogen lines, leading to a lower $\kappa_\mathrm{R}$ for $T > 15\,000\,\mathrm{K}$ \citep[e.g.,][]{Rogers2002,Mendoza2007}. 

Our atmospheric integrator as initially designed did not include metals, which resulted quite systematically in a nonconvergence for low-irradiated planets: the lack of dust grains made the atmosphere virtually transparent, and an intense, nonphysical surface gravity had to be set to counterbalance, via hydrostatic equilibrium, the very efficient internal radiation. Introducing trace quantities of metals (e.g., $Z = 0.0001$) has the benefit of restoring the atmospheric opacity without having to implement additional EoS, since the mean molecular weight stays almost unaltered. It should additionally be noted that the exact mixture of elements composing $Z$ is largely unknown in real-life cases, so in the \J{} code it is  set to match solar abundances \citep[according to][]{Asplund2021} by default. Even though sometimes host stars' compositions might be used as a proxy for the contents of their affiliated planets, such procedures remain hazardous because the correlation is not robust enough, especially for the smallest stars \citep[and references therein]{Hinkel2024}.

\section{The flagship case of GJ~436: Setting the scene}
\label{sect:preproc}

\subsection{Motivation}
\label{sect:motiv}

The aim of our previous work \citep{Attia2021} was to exhibit the potential of a HEM-delayed erosion scenario, enabled by our novel fully coupled approach. We sought to explore as many aspects as possible and explain the peculiarities of this system altogether. Here, the motivation is more quantitative: an exploration of the parameter space meant to predict the possible properties of the concealed companion, which would be responsible for a ZLK resonance that leads to a configuration compatible with the observed reality. In doing so, most of the use cases of Fig.~\ref{fig:jade_use_cases} are exploited, ultimately culminating in RE1 and RE2. Before diving into such an analysis, it is important to take a step back to examine our target and identify the questions to be answered. We refer to \citet{Attia2021} for a more comprehensive description of the system and we recapitulate here the main elements that make this tenant of the hot Neptune desert of particular interest in this paper.

\paragraph{An eccentric orbit.} GJ~436~b has been the subject of a growing interest in the past couple of decades, making it the target of several high-precision radial-velocity (RV) campaigns. One of the undisputed results of such Doppler measurements is the elliptic nature of its orbit \citep[eccentricity of $e = 0.152 \pm 0.009$,][]{Trifonov2018}, which has been progressively confirmed with higher and higher confidence through the years \citep{Butler2004,Butler2006,Maness2007,Knutson2014,Lanotte2014,Turner2016,Rosenthal2021}. The close proximity to the star \citep[semimajor axis of $a = 0.02858 \pm 0.00043\,\mathrm{AU}$,][]{Maxted2022} should have led tidal friction to circularize the orbit a long time ago, given the advanced age of the system \citep[$4 - 8\,\mathrm{Gyr}$,][]{Bourrier2018}. A straightforward solution could be the possible weakness of tidal forces \citep{Trilling2000,Mardling2008}. However, this argument is solely grounded in our poor knowledge of the tidal dissipation factor $Q_\mathrm{p}$, which is required to be abnormally high for this hypothesis to work. Alternatively, \citet{Beust2012} formulated a more convincing explanation based on ZLK cycles generated by a distant, unseen companion and ending in a late inward migration \citep[inspired by][]{Tong2009,Batygin2009}, which would also have the merit of solving the other unanswered interrogations presented below.

\paragraph{A polar architecture.} Motivated by the appealing HEM solution to the eccentricity conundrum, \citet{Bourrier2018} measured the spin--orbit angle of GJ~436~b to investigate if it validates this scenario. The result is crystal clear: the orbit is not only misaligned, but joins the striking sample of polar ones \citep{Albrecht2021,Attia2023,Siegel2023}, lending even more credence to HEM. Such a finding was later confirmed with more precise RV data and the advent of the advanced Rossiter--McLaughlin revolutions \citep{Bourrier2021} analysis technique \citep[3D spin--orbit angle of $\psi = 103^{+13}_{-12}{}^\circ$,][]{Bourrier2022}.

\paragraph{An evaporating atmosphere.} Thanks to spectroscopic transit observations in Lyman-$\alpha$, giant clouds of escaping hydrogen have been revealed enshrouding GJ~436~b and trailing the warm Neptune along its orbit \citep{Kulow2014,Ehrenreich2015,Lavie2017,dosSantos2019}. Subsequent models took profit of these results and unraveled many important characteristics of the eroding atmosphere, such as its comet-like structure \citep[e.g.,][]{Shaikhislamov2018,Khodachenko2019,VillarrealdAngelo2021}, the space weather around it \citep{Vidotto2017,Bellotti2023,Vidotto2023}, and, crucially its mass-loss rate \citep[$\sim 2.5 \times 10^8\,\mathrm{g/s}$,][]{Salz2016b,Bourrier2016}. While the latter value would require $\sim\,200\,\mathrm{Gyr}$ to erode just 1\% of the planet's total mass, the situation was vastly different during the star's youth. During the stellar saturation phase in the first few hundred Myr, XUV luminosities are typically $100 - 1000 \times$ higher than present values \citep{Jackson2012,Pezzotti2021}. Since atmospheric escape rates scale approximately linearly with XUV flux \citep[as seen in the energy-limited regime, e.g.,][]{Lammer2003}, this implies mass-loss timescales on the order of $200\,\mathrm{Myr}$ for each percent of total mass during the saturation phase. Over the system's lifetime ($> 4\,\mathrm{Gyr}$), this would result in several percent of cumulative mass loss: sufficient to substantially alter the planet's radius given that H/He atmospheres contribute disproportionately to planetary radii \citep[e.g.,][]{Zeng2019}. Such erosion could have pushed GJ~436~b well outside the hot Neptune desert boundaries. The HEM scenario thus reconciles the atmospheric escape we are witnessing today with the planet's survival at the inner fringes of the desert, explaining how it retains a substantial volatile envelope despite its advanced age.

\subsection{Bulk composition}
\label{sect:bulk}

A few key steps are required prior to any exploration of the parameter space (Appendix~\ref{app:jade_101}). Determining GJ~436~b's bulk structure and composition (RS1, RS2, Fig.~\ref{fig:jade_use_cases}) needs to be carried out first, since it conditions the outcome of any other subsequent analysis. The preferred approach would be a Bayesian inference of all the relevant parameters, constrained by today's observed properties \citep[e.g.,][]{Acuna2021}. Nevertheless, with only the planet mass and radius as observational inputs, the result is highly degenerate and no clear solution emerges. Instead, we chose to employ the following ``standard'' simplifying assumptions, which are not expected to heavily influence the secular evolution:

\begin{itemize}
    \item The mantle-to-core mass ratio is set to 2:1, similar to that of Earth;
    \item The atmospheric helium mass fraction is set to $Y = 0.2$, much like Neptune \citep{Hubbard1995,Helled2020};
    \item The atmospheric metal mass fraction is set to $Z = 10^{-5}$, a trace amount that still provides a dominant source of opacity at low temperatures (Fig.~\ref{fig:opacity_z}) without affecting the mean molecular weight (see discussion in Sect.~\ref{sect:jade_new}).
\end{itemize}

\noindent In this way, only one degree of freedom remains when it comes to the planet structure: $M_\mathrm{core}$, the core mass.\footnote{The word ``core'' has been used interchangeably in its astrophysical meaning, the part of the planet not consisting of the H/He atmosphere, and its geophysical meaning, the central part of the planet made of $\alpha$-Fe. Here, $M_\mathrm{core}$ is the mass of the former.} This parameter is derived by running a large set of internal structures under the above assumptions, with a fixed total mass of $M_\mathrm{p} = 21.68\,M_\oplus$ \citep{Maxted2022} and the present-day orbital properties for all of them. Each simulation differs from the others by its core mass, and the model that is ultimately retained is the one that yields the observational radius, $R_\mathrm{p} = 3.943 \pm 0.076\,R_\oplus$ \citep{Maxted2022}. All the planet structures were simulated at $t = 6\,\mathrm{Gyr}$, median of the system age range \citep{Bourrier2018}, as the time is critical in terms of controlling the stellar luminosity \citep[computed using \texttt{GENEC}, a stellar model adjusted to our system,][]{Eggenberger2008} and planetary internal heating; and, thus, in influencing the global radiative budget. Accordingly, we found that 89\% of today's planet mass is enclosed in the solid material, which corresponds to $M_\mathrm{core} = 19.30\,M_\oplus$. The core mass remains unchanged in any evolutionary simulation: we assume for simplicity that the iron and silicates do not dissolve in the volatile envelope; although it could be a possibility for such giant planets \citep{Guillot2004,Wilson2012}. Consequently, this invariable value of $M_\mathrm{core}$ is the one employed in all simulations of the exploration. The aforementioned assumptions are also imposed in the exploration, for a clear consistency.

\subsection{Grid of internal structures}
\label{sect:grid}

We note that a full exploration including hundreds of thousands of evolutionary models would be made unfeasible by the prohibitive computational cost of the structure integrator. Even for a single simulation, as the orbit evolves over billions of years, as the stellar XUV luminosity gradually declines, as the internal heating progressively fades out, and as the atmosphere evaporates, the planet's internal structure (and, ultimately, the planet radius) need to be regularly adjusted to follow these changes, which intensively calls the structure integrator. To overcome this limitation, a grid of internal structure models was precomputed, covering a maximum number of configurations that could arise during the exploration (FA2, Fig.~\ref{fig:jade_use_cases}). Once the ``static'' bulk parameters (RS1, RS2, Fig.~\ref{fig:jade_use_cases}) have been fixed, only three physical properties influencing the planet radius remain variable: the planet mass, $M_\mathrm{p}$, the equilibrium temperature, $T_\mathrm{eq}$ (related to stellar irradiation), and the intrinsic temperature, $T_\mathrm{int}$ \citep[related to planetary internal luminosity, with both temperatures being rigorously defined in][]{Attia2021}. The constructed grid spans 30 $M_\mathrm{p}$ values, 42 $T_\mathrm{eq}$ values, and 28 $T_\mathrm{int}$ values, yielding a total of $35\,280$ distinct points. The chosen values for each of the three parameters result from a uniform sampling of their conservatively large intervals of definition ($M_\mathrm{p} \in \left[M_\mathrm{core}; \, 80\,M_\oplus\right]$, $T_\mathrm{eq} \in \left[150\,\mathrm{K}; \, 1\,500\,\mathrm{K}\right]$, and $T_\mathrm{int} \in \left[10\,\mathrm{K}; \, 450\,\mathrm{K}\right]$) later refined where convergence is difficult (i.e., $M_\mathrm{p} \gtrsim M_\mathrm{core}$, and close to the definition boundaries for $T_\mathrm{eq}$ and $T_\mathrm{int}$).

The key parameter derived from every point of the grid is the planet radius, which is essential for the proper calculation of the dynamical forces (tides in particular) and the mass-loss rate. We could simply stop at the creation of this grid to complete FA2, by exporting the computed radii and interpolating them as a function of $M_\mathrm{p}$, $T_\mathrm{eq}$, and $T_\mathrm{int}$ within the evolutionary simulations that are part of the exploration, which would still drastically decrease the run time in theory. However, a few points in the grid actually fail to converge, especially in the regions with a refined resolution, breaking down the grid-like structure. If interpolation is relatively straightforward on a regular mesh (via, e.g., trilinear interpolation), schemes defined for scattered points are much more complex, in particular, for a dimensionality higher or equal than three. In our case, numerical methods are not guaranteed to converge and would sometimes negate the expected run time gain in practice.

\subsection{Mass--radius relationships}
\label{sect:mr}

\begin{figure}
\includegraphics[width=\linewidth]{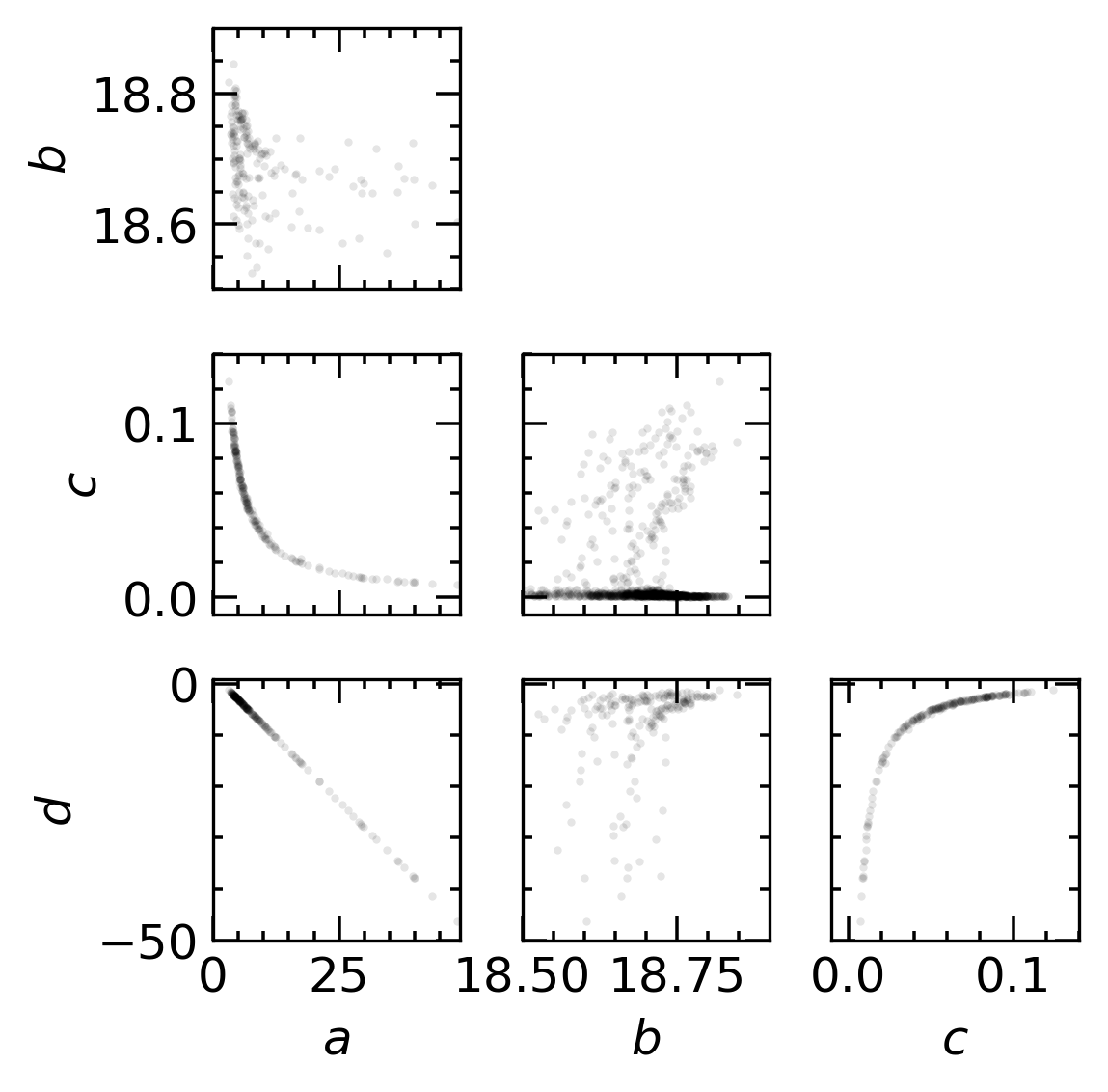}
\centering
\caption{Corner plot for the fitted coefficients $a$, $b$, $c$, and $d$ from Eq.~(\ref{eq:MR}). Each black point is the result of fitting one internal structure model from the precomputed grid (Sect.~\ref{sect:grid}).}
\label{fig:gj436_coeff_corr}
\end{figure}

\begin{table}
\caption{Fitting coefficients for Eqs.~(\ref{eq:MR_c}) and (\ref{eq:MR_d}).}
\centering
\begin{tabular}{c c}
\hline
Coefficient & Value \\
\hline
\hline
$c_0$ & -1.0000199845566071 \\
$c_1$ & -1.0000166831011739 \\
$d_0$ & -0.9999953510919015 \\
$d_1$ & 1.9998974861128505 \\
\hline
\end{tabular}
\label{tab:coeff_fit}
\end{table}

Instead, we derive analytical relationships for $R_\mathrm{p}$ as a function of the other parameters. Evaluating a scalar function (instead of interpolating) is the optimal solution from a numerical standpoint, but it requires ``guessing'' the shape the formula would take. We could, for example, try to derive it from first principles \citep[as ingenuously executed by][]{Turbet2020} and then compare its results to the grid radii. A simpler and more flexible solution is to utilize a generic parametric form, subsequently adjusted to fit the data \citep[much like the vast majority of mass--radius relationships in the literature, e.g.,][]{Grasset2009,Swift2012,Zeng2016,Otegi2020,Parc2024}, which is the approach chosen here. Attempts at constructing a radius function with an explicit dependency to all three parameters $M_\mathrm{p}$, $T_\mathrm{eq}$, and $T_\mathrm{int}$ unfortunately gave inconclusive results. Instead, we resorted to classical mass--radius relationships, whose coefficients depend on the two temperatures. The simplest form we might conceive of would be a power law $R_\mathrm{p}\,\propto\,M_\mathrm{p}^a$, where $a$ is the power-law exponent. An observational fit to the telluric planets of the Solar System yields $a = 0.33$ \citep[e.g.,][]{Wolfgang2012} like for constant density spheres, whereas including gas giants gives higher values \citep[e.g.,][]{Lissauer2011}, but the most massive super-Jupiters are expected to have a negative exponent \citep{Chabrier2009}. To account for these different regimes, the best parametric form modulating the power law, which we found by trial and error, being initially inspired by \citet{Mordasini2012,Aguichine2021}, is the following

\begin{equation}
\label{eq:MR}
\log R_\mathrm{p} = a \left(\log \frac{M_\mathrm{p}}{b}\right)^c + d,
\end{equation}

\noindent where $a$, $b$, $c$, and $d$ are coefficients to be fitted. In this regard, they are determined for each isotherm $\left(T_\mathrm{eq}, \, T_\mathrm{int}\right)$ in the grid by a least-squares minimization, so that Eq.~(\ref{eq:MR}) delivers the closest radius possible to the one computed by the \J{} code.\footnote{The numerical values of the parametric coefficients are all derived assuming $M_\mathrm{p}$ and $R_\mathrm{p}$ are expressed in units of $M_\oplus$ and $R_\oplus$ respectively, and the temperatures in K. Natural logarithms are employed in Eqs.~(\ref{eq:MR}) and (\ref{eq:MR_c}).} Importantly, scrutinizing the results in the form of a pair-by-pair corner plot (Fig.~\ref{fig:gj436_coeff_corr}) reveals that some coefficients are in fact correlated. This is a considerable advantage since it allows us to reduce the number of free parameters: $c$ and $d$ can be expressed as a function of $a$, a power law for the former and a simple linear relation for the latter, as explicitly seen in Fig.~\ref{fig:gj436_coeff_corr}. Therefore, we can write

\begin{equation}
\label{eq:MR_c}
\log c = c_0 \log a + c_1,
\end{equation}

\begin{equation}
\label{eq:MR_d}
d = d_0 a + d_1,
\end{equation}

\noindent where $c_0$, $c_1$, $d_0$, and $d_1$ are, this time, unique coefficients (constant for any combination of $\left(T_\mathrm{eq}, \, T_\mathrm{int}\right)$). Their exact values, also found with a least-squares minimization, are shown in Table~\ref{tab:coeff_fit}, but strongly suggest that $c\,\propto\,1/a$ and $d \simeq 2 - a$. Their extreme proximity to integer values might be indicative of a physical reason, but we did not find any convincing lead. As a matter of fact, fixing them to their integer counterparts degrades the goodness of the fit, so we decided to use the values of Table~\ref{tab:coeff_fit} as they are.

\begin{figure*}
\includegraphics[width=\linewidth]{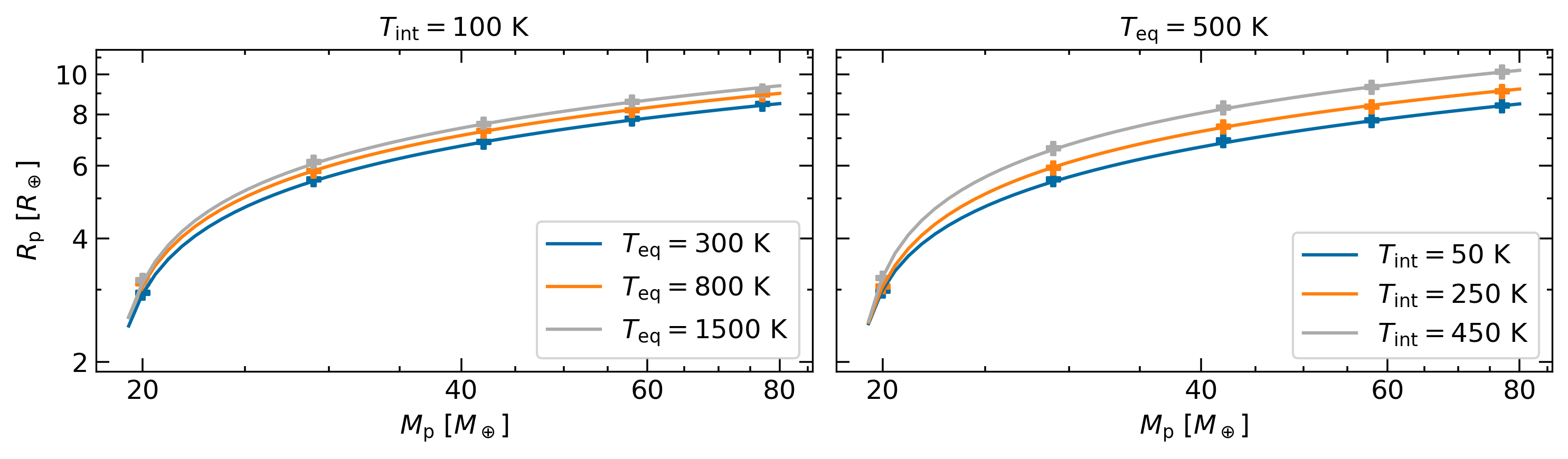}
\centering
\caption{Analytical mass--radius relationships for GJ~436~b. The crosses represent some radii directly computed by the structure integrator to emphasize the goodness of the fit.
\textbf{Left:} Varying equilibrium temperature, $T_\mathrm{eq}$, with a fixed intrinsic temperature $T_\mathrm{int} = 100\,\mathrm{K}$.
\textbf{Right:} Varying intrinsic temperature $T_\mathrm{int}$, with a fixed equilibrium temperature $T_\mathrm{eq} = 500\,\mathrm{K}$.}
\label{fig:gj436_mr}
\end{figure*}

To summarize, the procedure to calculate the planet radius at any moment of any evolutionary simulation can be sketched this way:

\begin{enumerate}
    \item Computing $T_\mathrm{eq}$ and $T_\mathrm{int}$ \citep{Attia2021};
    \item Bilinearly interpolating $a$ and $b$ from $T_\mathrm{eq}$ and $T_\mathrm{int}$;
    \item Computing $c$ and $d$ from $a$ using Eqs.~(\ref{eq:MR_c}) and (\ref{eq:MR_d});
    \item Injecting $M_\mathrm{p}$, $a$, $b$, $c$, and $d$ in Eq.~(\ref{eq:MR}).
\end{enumerate}

\noindent Compared to one run of the structure integrator, determining the radius this way represents an immense run time gain, as illustrated in Appendix~\ref{app:perf}. Figure~\ref{fig:gj436_mr} depicts some mass--radius relationships for a couple of combinations of $T_\mathrm{eq}$ and $T_\mathrm{int}$. On each isotherm are overplotted a few radii directly computed by \J{}'s structure integrator, showcasing the agreement with the analytical formulae, as also further explored in Appendix~\ref{app:perf}.

\section{The flagship case of GJ~436: Exploration}
\label{sect:explo}

\subsection{Framework}
\label{sect:fw}

The aim of this investigation is to constrain at the same time the initial conditions, after the disk dissipation, of planet b (RE1, Fig.~\ref{fig:jade_use_cases}), as well as the properties of the hypothetical planet c (RE2, Fig.~\ref{fig:jade_use_cases}). To this effect, we employed a novel framework that combines grid-based forward modeling with Bayesian inference. We termed this approach ``semi-Bayesian'' because even though we did use Bayesian methods (priors, likelihoods, and MCMC sampling) to explore the parameter space, the likelihood computation relies on precomputed grids of simulations rather than on-the-fly calculations. Specifically, we first generated a comprehensive grid of fully coupled simulations carried out over the upper age limit of the system, $8\,\mathrm{Gyr}$, spanning a wide domain of the parameters to be constrained. The likelihood at any point in parameter space was then obtained by interpolating between these precomputed grid points.

When it comes to RE1, the two main properties that we explore, characterizing the infancy of GJ~436~b, are the initial semimajor axis, $a_0$, and the initial planet mass, $M_\mathrm{p,\,0}$. The initial eccentricity, argument of periastron, and spin--orbit angle are left unexplored because, provided that the ZLK mechanism occurs, they are bound to oscillate. Their initial values, unless they are extreme, will be regularly reached during the long-term secular evolution of the system, because the ZLK characteristic timescale, $\tau_\mathrm{ZLK}$, imposing the period of the cycles within a factor on the order of unity \citep{Beust2006}, is much smaller than the global timescale of the evolution. In other words, two systems with different initial eccentricities (or arguments of periastron and spin--orbit angles), all else being equal, will differ by a phase shift. While the previous statement is technically not entirely true, as very specific and well-chosen initial conditions can for instance restrict the amplitude of the cycles, we performed a few such simulations to corroborate its conclusion. In this way, all inner orbits were initialized as circular and aligned with the spin of the star. The planet axial tilt, monitored by the \J{} code, also starts as null.

As for RE2, the perturber would be fully defined, ZLK-wise, via its mass, $M_\mathrm{pert}$, semimajor axis, $a_\mathrm{pert}$, eccentricity, $e_\mathrm{pert}$, and initial mutual inclination, $i_\mathrm{mut,\,0}$, with respect to the inner orbit. We assume for simplicity to truncate the Hamiltonian up to the quadrupolar order, which enables very fast computations without an excessive accuracy loss. \citet{Beust2012} carried out dynamical studies at various orders, showing that the two-fold ZLK evolution reported below (Fig.~\ref{fig:gj436_maxlike}) still holds whatever the truncation order. Only the transition time, $t_\mathrm{trans}$, appears sensitive to the truncation order (and defined in Sect.~\ref{sect:resfid}), with an overestimation of $\sim\,30\%$ at the quadrupolar level \citep[Fig.~4 of][]{Beust2012}. Within this hypothesis, we deduce from the equations of motion \citep[Appendix~A.3 of][]{Attia2021} that the mass, semimajor axis, and eccentricity of the companion can in fact be combined into a single parameter controlling the intensity of the resonance,

\begin{equation}
\label{eq:Lambda}
\Lambda \equiv \frac{a_\mathrm{pert}^3 \left(1 - e_\mathrm{pert}^2\right)^{3/2}}{\mathcal{G}M_\mathrm{pert}},
\end{equation}

\noindent which is the parameter we explore in the end, allowing for a very welcome reduction in dimensionality. In all occurrences where numerical values of $\Lambda$ are given, $a_\mathrm{pert}$ is assumed to be in AU, $M_\mathrm{pert}$ in $M_\oplus$, and the gravitational constant, $\mathcal{G} = 1$. Of course, we numerically checked that two perturbers with different properties leading to the same value of $\Lambda$ actually generated an identical ZLK resonance. It should be noted that $\Lambda$ is proportional to a time squared, and is highly reminiscent of the ZLK characteristic timescale. This is no coincidence, and the two quantities can in fact be linked through $\tau_\mathrm{ZLK} \simeq \Lambda/P$, with $P$ as the inner orbital period. All in all, three parameters needed to be constrained: $a_0$, $M_\mathrm{p,\,0}$, and $\Lambda$. For this specific section and the following, the initial mutual inclination is fixed to $i_\mathrm{mut,\,0} = 75^\circ$ as a typical value so as to isolate its influence (Sect.~\ref{sect:resfid}), but it was also explored in a second time (Sect.~\ref{sect:incl}).

A fundamental challenge arises when comparing our evolutionary models to observations regarding how we can evaluate the compatibility between simulations that predict time-varying planetary properties over $8\,\mathrm{Gyr}$ and observational data that represent a snapshot of the system at a single, uncertain epoch. The system's age is only constrained to lie within $4 - 8\,\mathrm{Gyr}$ and properties such as eccentricity, mass, and spin--orbit angle all evolve throughout this window. Simply checking whether a simulation matches observations at a fixed time would be inadequate, as it would miss potentially valid evolutionary pathways that reproduce the observed state at different epochs within the age uncertainty. Therefore, we would need a statistical framework that accounts for both the temporal evolution of the simulated properties and the uncertainty involved when these properties would be match up with the observations.

Then, for each 8-Gyr fully joint simulation corresponding to a distinct point $\left(a_0, \, M_\mathrm{p,\,0}, \, \Lambda\right)$ in the explored 3D parameter space, a score function needs to be evaluated to quantify the compatibility with the currently observed parameters. To this effect, each simulation can be broken down into a temporal collection of evolving system properties (time, semimajor axis, eccentricity, mass, etc.). For every point $i$ of the time series, a log-likelihood of the constraining data (today's observed parameters) with respect to the model $\log p_i$ is computed using the distance between $i$ and the posterior distribution underlying the constraining data. To make things clearer, we can take a concrete example, with only one constraining parameter: today's eccentricity, with its error bar, $e_\bullet \pm \sigma\left(e_\bullet\right)$. Observational measurements are routinely assumed to take the shape of a Gaussian distribution, the log-distance between $i$ and the constraining eccentricity would be $-\left((e_i - e_\bullet)/\sigma\left(e_\bullet\right)\right)^2/2$, where $e_i$ is the eccentricity of point $i$ of the simulation. The latter quantity needs to be complemented with a time information, since obtaining the correct eccentricity needs to happen within the correct time span as well. Here, the constraint we have on the system age is uniform between two bounds, $t_{\bullet\downarrow}$ and $t_{\bullet\uparrow}$, which gives the total log-likelihood for point $i$ of the simulation constrained by the current eccentricity,

\begin{equation}
\label{eq:logPi}
\log p_i = -\frac{1}{2}\left(\frac{e_i - e_\bullet}{\sigma(e_\bullet)}\right)^2 + 
            \log\left(\mathds{1}_{[t_{\bullet\downarrow},\,t_{\bullet\uparrow}]}(t_i)\right).
\end{equation}

\noindent This value can be obtained up to an additive constant, with $t_i$ the time value of point $i$ and $\mathds{1}$ the indicator function that is equal to 1 when $t_i$ falls within the allowed age range and 0 otherwise. This ensures that only configurations matching the observations within the system's age constraints contribute to the total likelihood, with no preference for any particular age within that range. We can see from Eq.~(\ref{eq:logPi}) that the $i$ points maximizing their $\log p_i$ are the ones yielding the closest $e_i$ to $e_\bullet$ during the time interval $\left[t_{\bullet\downarrow}, \, t_{\bullet\uparrow}\right]$, which is exactly the desired behavior. In practice, there are several constraining observational measurements; each contribution is added to $\log p_i$, assuming it is independent from the others. Finally, the total log-likelihood of a $\left(a_0, \, M_\mathrm{p,\,0}, \, \Lambda\right)$ simulation is computed by numerically time-integrating the various $p_i$ pertaining to the individual points of said simulation,

\begin{equation}
\label{eq:like}
p = \int p_i(t) \mathrm{d}t \simeq \sum_i p_i \Delta t_i.
\end{equation}

\noindent From this, we can then calculate

\begin{equation}
\label{eq:logP}
\log p = \log p_\mathrm{max} + \log \sum_i \exp\left(\log p_i - \log p_\mathrm{max}\right) \Delta t_i,
\end{equation}

\noindent with $\log p_\mathrm{max}$ as the maximal value of the different $\log p_i$, this formulation of Eq.~(\ref{eq:logP}) avoiding null values at machine precision in the evaluation of the $\exp$ function. It is important to note that this temporal integration approach inherently addresses the likelihood of observing the system in its current state. Simulations that match the observational constraints for longer durations within the allowed age range $\left[t_{\bullet\downarrow}, \, t_{\bullet\uparrow}\right]$ naturally accumulate higher total log-likelihoods through the summation in Eq.~(\ref{eq:logP}). Conversely, configurations that only briefly pass through the observed parameter space, such as those undergoing rapid migration, receive proportionally lower scores. This weighting scheme thus automatically penalizes fine-tuned solutions that require precise timings to match current observations. We describe more rigorously the construction of the probability function in Appendix~\ref{app:bayes}.

After setting a uninformative prior for all three jump parameters, we were equipped with a grid of \J{} simulations, paving the $\left(a_0, \, M_\mathrm{p,\,0}, \, \Lambda\right)$ parameter space, each one of them associated with a $\log p$. Intermediate values, uncovered by \J{} models, have their $\log p$ interpolated from the grid. Ultimately, the regions of high likelihood are sought for by employing an MCMC implementation using \texttt{emcee} \citep{ForemanMackey2013}. Because such regions can be highly multimodal, we found that using a custom ``move'' parametrization was best to efficiently explore the entirety of the parameter space (a move is the algorithm for proposing an update of the coordinates of the sampled posterior at each step). In our case, a well-behaved setup is a weighted combination of a differential evolution proposal \citep{Nelson2013} with a probability of 90\% and a snooker move \citep{terBraak2008} with a probability of 10\%, which has the crucial benefit of avoiding being stranded in a local maximum by occasionally allowing for large moves. In any case, the length of the MCMC chains (typically $10\,000$), number of walkers (typically 64), and burn-in phase (typically $5\,000$) are adjusted to ensure convergence. \\

\begin{table}
\caption{Fixed parameters for the \J{} simulations of the fiducial exploration (Sect.~\ref{sect:resfid}).}
\centering
\setlength\tabcolsep{0pt}
\begin{tabular*}{\linewidth}{@{\extracolsep{\fill}} cccccc }
\hline
Parameter & Symbol & Value \\
\hline
\hline
Total simulation time & $t_\mathrm{f}$ & $8\,\mathrm{Gyr}^\mathrm{(a)}$ \\
Initial simulation time & $t_0$ & $10\,\mathrm{Myr}^\mathrm{(b)}$ \\
\hline
Stellar mass & $M_\star$ & $0.445\,M_\odot^\mathrm{(c)}$ \\
Stellar radius & $R_\star$ & $0.425\,R_\odot^\mathrm{(c)}$ \\
Stellar apsidal constant & $k_\star$ & $0.01^\mathrm{(d)}$ \\
Stellar tidal dissipation constant & $Q_\star$ & $10^5{}^\mathrm{(d)}$ \\
Stellar moment of inertia parameter & $\alpha_\star$ & $0.08^\mathrm{(d)}$ \\
Initial stellar spin rate & $\Omega_{\star,\,0}$ & $280\,\mathrm{yr}^{-1}{}^\mathrm{(e)}$ \\
\hline
Inner planet core mass & $M_\mathrm{core}$ & $19.30\,M_\oplus^\mathrm{(f)}$ \\
Inner planet apsidal constant & $k_\mathrm{p}$ & $0.25^\mathrm{(d)}$ \\
Inner planet tidal dissipation constant & $Q_\mathrm{p}$ & $10^5{}^\mathrm{(d)}$ \\
Inner planet moment of inertia parameter & $\alpha_\mathrm{p}$ & $0.25^\mathrm{(d)}$ \\
Inner planet atmospheric helium fraction & $Y$ & $0.2^\mathrm{(f)}$ \\
Inner planet atmospheric metal fraction & $Z$ & $10^{-5}{}^\mathrm{(f)}$ \\
Inner planet rock-to-iron fraction & $f_\mathrm{Si/Fe}$ & $2^\mathrm{(f)}$ \\
\hline
Inner planet initial eccentricity & $e_0$ & $10^{-4}$ \\
Inner planet initial argument of periastron & $\omega_0$ & $0^\circ$ \\
Inner planet initial spin--orbit angle & $\psi_0$ & $0^\circ$ \\
Inner planet initial axial tilt & $\varphi_0$ & $0^\circ$ \\
Outer planet semimajor axis & $a_\mathrm{pert}$ & $20\,\mathrm{AU}$ \\
Outer planet eccentricity & $e_\mathrm{pert}$ & $10^{-4}$ \\
Outer planet argument of periastron & $\omega_\mathrm{pert}$ & $0^\circ$ \\
Initial mutual inclination & $i_\mathrm{mut,\,0}$ & $75^\circ$ \\
\hline
\end{tabular*}
\begin{tablenotes}
$^\mathrm{(a)}$~system age upper limit \citep{Bourrier2018}~;
$^\mathrm{(b)}$~typical disk dispersal time~;
$^\mathrm{(c)}$ \citet{Maxted2022}~; $^\mathrm{(d)}$ \citet{Attia2021}~;
$^\mathrm{(e)}$~stellar model \citep{Eggenberger2008} adjusted to the system~;
$^\mathrm{(f)}$~Sect.~\ref{sect:bulk}.
\end{tablenotes}
\label{tab:fixed_param}
\end{table}

\subsection{Results: Fiducial inclination}
\label{sect:resfid}

\begin{table*}
\caption{Constraints for the \J{} simulations of the exploration.}
\centering
\setlength\tabcolsep{0pt}
\begin{tabular*}{\linewidth}{@{\extracolsep{\fill}} ccc }
\hline
Parameter & Symbol & Constraint \\
\hline
\hline
Time & $t$ & $\mathcal{U}\left(t_{\bullet\downarrow} = 4\,\mathrm{Gyr};\,t_{\bullet\uparrow} = 8\,\mathrm{Gyr}\right)^\mathrm{(a)}$ \\
Inner planet semimajor axis & $a$ & $\mathcal{G}\left(a_\bullet = 0.02858\,\mathrm{AU};\,\sigma(a_\bullet) = 0.00043\,\mathrm{AU}\right)^\mathrm{(b)}$ \\
Inner planet eccentricity & $e$ & $\mathcal{G}\left(e_\bullet = 0.152;\,\sigma(e_\bullet) = 0.009\right)^\mathrm{(c)}$ \\
Inner planet spin--orbit angle & $\psi$ & $\mathcal{G}\left(\psi_\bullet = 103^\circ;\,\sigma(\psi_\bullet) = 13^\circ\right)^\mathrm{(d)}$ \\
Inner planet mass & $M_\mathrm{p}$ & $\mathcal{G}\left(M_\mathrm{p\bullet} = 21.68\,M_\oplus;\,\sigma(M_\mathrm{p\bullet}) = 0.63\,M_\oplus\right)^\mathrm{(b)}$ \\
Inner planet radius & $R_\mathrm{p}$ & $\mathcal{G}\left(R_\mathrm{p\bullet} = 3.943\,R_\oplus;\,\sigma(R_\mathrm{p\bullet}) = 0.076\,R_\oplus\right)^\mathrm{(b)}$ \\
\hline
\end{tabular*}
\begin{tablenotes}
$\mathcal{U}\left(x_\downarrow;\,x_\uparrow\right)$ is a uniform distribution in $\left[x_\downarrow;\,x_\uparrow\right]$.
$\mathcal{G}\left(x;\,\sigma(x)\right)$ is a Gaussian distribution centered on $x$ with a standard deviation of $\sigma(x)$. 
$^\mathrm{(a)}$ \citet{Bourrier2018}~; $^\mathrm{(b)}$ \citet{Maxted2022}~; $^\mathrm{(c)}$ \citet{Trifonov2018}~; $^\mathrm{(d)}$ \citet{Bourrier2022}.
\end{tablenotes}
\label{tab:constraints}
\end{table*}

To maintain the readability of the results, we fixed the initial mutual inclination to a typical value $i_\mathrm{mut,\,0} = 75^\circ$ as a first step. In total, the base grid of simulations we carried out for this fiducial exploration comprises $113\,986$ points, each corresponding to a different $\left(a_0, \, M_\mathrm{p,\,0}, \, \Lambda\right)$ configuration. Table~\ref{tab:fixed_param} shows the set of fixed parameters common to all simulations. When it comes to the companion properties, we imposed a circular orbit at $20\,\mathrm{AU}$. This way, the explored values of $\Lambda$ are unambiguously controlled by $M_\mathrm{pert}$ via Eq.~(\ref{eq:Lambda}), preserving in all cases the hierarchical triple assumption ($a_\mathrm{pert} \gg a$). This does not mean that the companion lies in fact at $20\,\mathrm{AU}$, as a $\Lambda$ solution is degenerate between an infinity of compatible combinations of $a_\mathrm{pert}$, $e_\mathrm{pert}$, and $M_\mathrm{pert}$. We recall that the perturber remains unchanged during the entire simulation, acting like an angular momentum reservoir. Moreover, the inner planet's spin rate is always initialized to be synchronized with the orbital motion, as its value is unknown. As a matter of fact, it is a safe assumption because the planet spin has a minimal impact on the secular evolution in our situation, although it could have a greater influence on energy dissipation in a more general case, as part of advanced tidal models \citep{Makarov2013}. Plus, \citet{Beust2012} showed that during the resonance phase, the planet's spin tends to accelerate to synchronize with the orbital angular velocity at periastron in the high eccentricity phases of ZLK cycles, that is to say when SRFs are strongest. The choice of planet b's initial argument of periastron is motivated by the fact that $\omega = 0^\circ$ represents the locus of minimum eccentricity within a quadrupole approximation \citep{Lithwick2011}, consistent with the initial circular orbit. 

\begin{figure*}
\includegraphics[width=\linewidth]{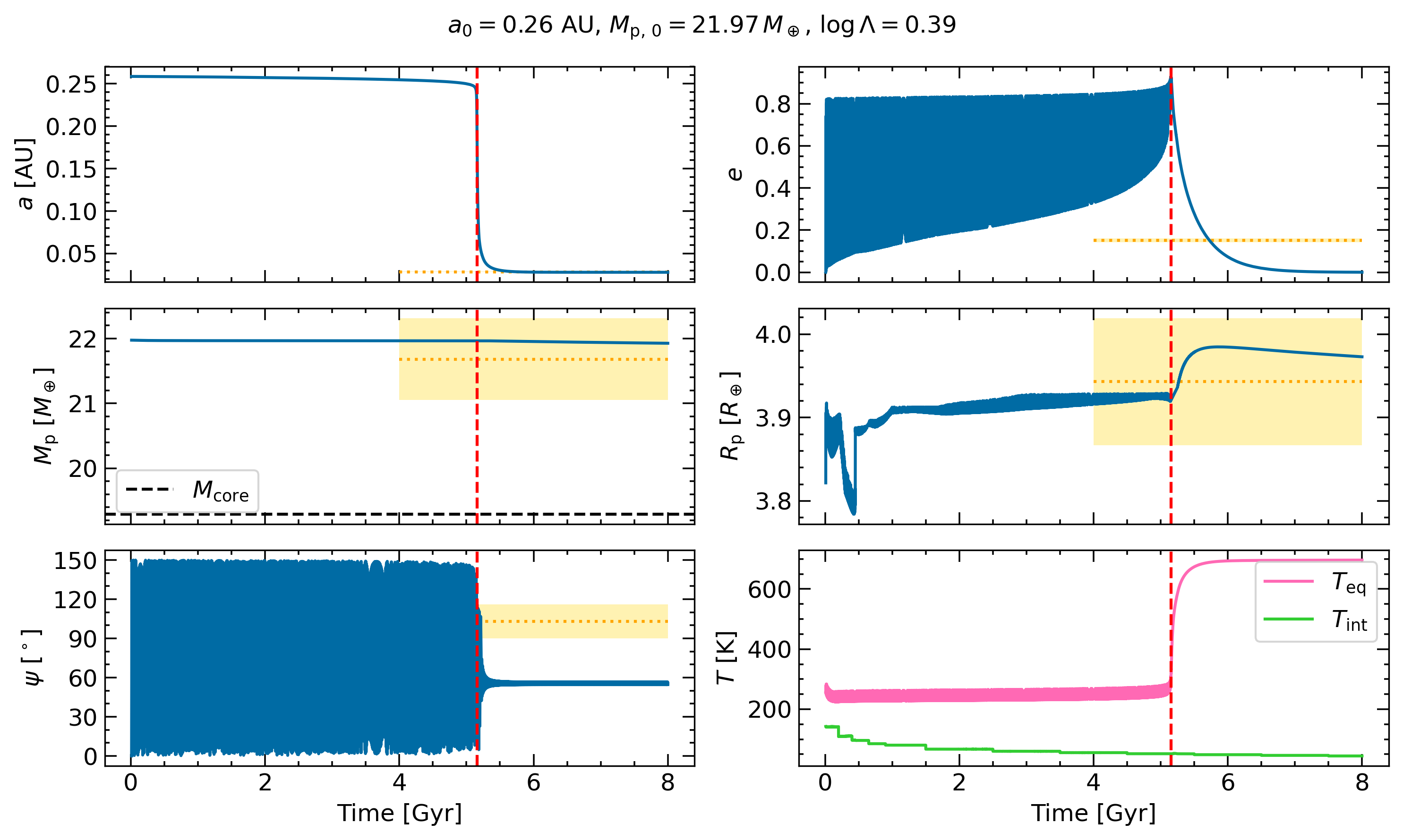}
\centering
\caption{Secular evolution of the system corresponding to the maximum log-likelihood of the fiducial exploration ($i_\mathrm{mut,\,0} = 75^\circ$, Sect.~\ref{sect:resfid}), as simulated by \J{}. The orange shaded regions represent the observational constraints of Table~\ref{tab:constraints}. In all plots, $t_\mathrm{trans}$ is shown as a dashed vertical red line.}
\label{fig:gj436_maxlike}
\end{figure*}

Table~\ref{tab:constraints} lists the various observational measurements constraining the evolutionary simulations and contributing to the calculation of $\log p$ (Eq.~(\ref{eq:logP})). As seen in Fig.~\ref{fig:gj436_corner}, a clear solution emerges, embodied by the peaks of maximum likelihood in the 1D histograms. Configurations falling within the shaded regions, which can be interpreted as the uncertainty intervals for the jump parameters $\left(a_0, \, M_\mathrm{p,\,0}, \, \Lambda\right)$, yield outcomes satisfactorily close to the currently observed reality. This is best visible in Fig.~\ref{fig:gj436_maxlike}, simulating the evolution of a system corresponding to the medians of the posterior distributions (their numerical values are reported on top of the figure, as well as above the 1D histograms in Fig.~\ref{fig:gj436_corner}). All the observational constraints of Table~\ref{tab:constraints} are matched for this maximum-likelihood simulation, except for $\psi_\bullet$ as discussed later. The typical two-fold ZLK evolution is explicit. In a first phase, dominated by the resonance, large eccentricity oscillations are seen, modulated by a shrinking envelope due to SRFs. The spin--orbit angle wildly oscillates between $0^\circ$ and $150^\circ$, while the semimajor axis remains roughly constant. When the amplitude of the eccentricity cycles becomes negligible, GJ~436~b decouples from the companion: the resonance is over. This point in time $t_\mathrm{trans}$ is crucial as it marks the transition to the second phase, which is dominated by tidal damping, featuring an abrupt inward migration and locking the planet on a misaligned orbit. 

The total impact of mass loss is extremely mild, to say the least, but the radius variations are worth commenting. The sharp decrease at $t \sim\,200\,\mathrm{Myr}$ coincides with the end of the stellar saturation phase, when the XUV flux suddenly drops by orders of magnitude, leading to rapid atmospheric contraction \citep[similar behavior is shown in Figures~$9 - 11$ of][]{Attia2021}. The subsequent radius increase appears linked to $T_\mathrm{int}$ dropping below $100\,\mathrm{K}$ due to the sharp decrease in planetary luminosity at this epoch. While the exact mechanism causing this inflation at this specific temperature threshold remains unclear, it likely reflects a transition in the atmospheric opacity regime or convective properties that our mass--radius relationships capture empirically but whose physical origin merits further investigation. Then, the radius globally mimics the steady increase in the equilibrium temperature due to the slowly increasing mean eccentricity, before inflating at $t_\mathrm{trans}$ because of migration and the resulting additional heating. Finally, the atmosphere slowly contracts because of evaporation, despite it being minimal. The same radius pulsations in tune with the ZLK cycles, identified in \citet{Attia2021}, are reported all along the first phase.

Globally speaking, the following conclusions can be drawn from a meticulous inspection of the corner plot (Fig.~\ref{fig:gj436_corner}) and the underlying simulations. First, configurations from inside the uncertainty margins of the posterior distributions are generally compatible with Table~\ref{tab:constraints} within the error bars of the constraints (except perhaps for $\psi_\bullet$, see discussion below), which is exactly the coveted behavior, lending credence to our semi-Bayesian model.

The observational constraint on the semimajor axis is the most restrictive especially for the inferred $a_0$, which can be intuited from a theoretical standpoint. The value of the final separation $a_\mathrm{f}$ of GJ~436~b is controlled by tidal damping, main dynamical engine after the decoupling from the companion. During this phase, $a\left(1 - e^2\right)$ is a roughly constant\footnote{There is actually an angular momentum exchange between the orbit and the stellar spin, but the argument still holds at the order of magnitude level.} quantity thanks to the conservation of angular momentum \citep{Socrates2012}. Since the fate of the orbit is ultimately to circularize, the final separation is dictated by $a_\mathrm{f} = a(t = t_\mathrm{trans})\left(1 - e(t = t_\mathrm{trans})^2\right)$. Nonetheless, $a(t = t_\mathrm{trans}) \simeq a_0$ because tides, responsible for the orbit shrinkage, are largely surpassed by the resonance counteracting them during the first phase (as seen in Fig.~\ref{fig:gj436_maxlike}). Furthermore, $e(t = t_\mathrm{trans})$ can be seen as a proxy for the maximum eccentricity reached during the ZLK cycles $e_\mathrm{max}$, which only depends on the initial inner eccentricity/argument of periastron and the initial mutual inclination to first order within the quadrupole approximation \citep{Katz2011}. All these parameters being fixed (Table~\ref{tab:fixed_param}), $e_\mathrm{max}$ should be roughly the same for all explored configurations (limited variations might subsist due to SRFs). Consequently, $a_\bullet$ (the desired value for $a_\mathrm{f}$) constrains $a_0$ very strongly, in particular given the low uncertainty $\sigma(a_\bullet)$ due to the minute precision of CHEOPS \citep{Benz2021}. This argument is validated by the very peaked distribution of maximum eccentricities in the simulations where the resonance was triggered $e_\mathrm{max} = 0.941 \pm 0.005$, and the ensuing theoretical expectation $a_0 = a_\bullet/\left(1 - e_\mathrm{max}^2\right) = 0.249 \pm 0.009\,\mathrm{AU}$, which is satisfyingly consistent with the inferred value from the exploration (Fig.~\ref{fig:gj436_corner}).

The observational constraint on the eccentricity is actually not as restrictive as anticipated. It is understandable, \textit{a posteriori}, because $e_\bullet$ represents but a certain nonzero value that is eventually browsed after the decoupling, if $t_\mathrm{trans}$ falls reasonably within $\left[t_{\bullet\downarrow};\,t_{\bullet\uparrow}\right]$. This interval being quite large, the constraint brought by $e_\bullet$ on $t_\mathrm{trans}$ translates into a moderate constraint on $\Lambda$, which explains the variation of the latter of more than half a dex within $1\,\sigma$ of its posterior.

Similarly to the case of the maximum-likelihood simulation (Fig.~\ref{fig:gj436_maxlike}), evaporation has a minimal impact on compatible configurations in general. By forming farther out than its current orbit, GJ~436~b is spared from the infancy of the star, when it is most energetic \citep[e.g.,][]{Jackson2012}. When it migrates to its final position, it is already too late for evaporation to make a difference: the stellar XUV luminosity has already dropped by several orders of magnitude. In fact, the inferred $M_\mathrm{p,\,0}$ along with its uncertainty are encompassed in the error bar of $M_\mathrm{p \bullet}$, consistent with a null net effect of mass loss. Nevertheless, accurately modeling the atmospheric structure is still pivotal because of the radius pulsations and their meaningful influence on tides \citep[$\propto\,R_\mathrm{p}^5$, e.g.,][]{Jackson2008}, to which $t_\mathrm{trans}$ is extremely sensitive \citep{Beust2012,Attia2021}.

The observational constraint on the spin--orbit angle is the most difficult to meet. Only a handful of simulated planets emerge from the resonance with a compatible $\psi$. We can check these results by examining the distribution of final spin--orbit angles, $\psi_\mathrm{f}$, within simulations that have the expected two-fold behavior (Fig.~\ref{fig:gj436_psif}). The bulk of the orbits finish their runs on moderately tilted orbits between $15^\circ$ and $60^\circ$, and a tiny fraction ends up on a retrograde orbit. The most striking feature is the total absence of intermediate values of $\psi_\mathrm{f}$ between these two modes, which is related to our choice of initial mutual inclination. This final distribution is similar to the one produced by \citet{Fabrycky2007}, with the major difference that their two modes spread enough to overlap, leaving no void between them. This is due to the fact that they also investigated a large number of $i_\mathrm{mut,\,0}$. In any case, the role of $\psi_\bullet$ will become central when $i_\mathrm{mut,\,0}$ is explored (Sect.~\ref{sect:incl}).

The radius variations observed in our simulations warrant comparison with previous studies of coupled planetary evolution. While several works have investigated the effects of a dynamically evolving atmospheric structure, the specific radius pulsations we observe during ZLK cycles arise from a distinct mechanism. \citet{Gu2003} and \citet{Miller2009} focus on tidal heating's impact on planetary radii, a process not implemented in \J{}. However, since tidal heating scales with $\lvert\dot{a}/a\rvert$ \citep[neglecting spin variation,][]{Eggleton1998,Mardling2002} and the semimajor axis remains roughly constant during the ZLK phase (Fig.~\ref{fig:gj436_maxlike}), this omission should minimally affect our radius pulsations, which instead originate from fluctuations in stellar irradiation. \citet{Lopez2016} include all relevant contributions to the planet energy budget except tides (similar to \J{}, even though we do not implement them in a self-consistent way) but do not secularize the irradiation, focusing instead on radius evolution during the stellar postmain sequence. Our pulsations are fundamentally an eccentricity-driven phenomenon emerging from the secularization of irradiation power, which increases substantially as $e$ approaches unity during ZLK cycles \citep{Attia2021}. \citet{Glanz2022} implement a comprehensive treatment including secularized irradiation and could in principle capture these pulsations, though their focus is on the tidal circularization phase. We acknowledge that during the rapid migration after $t_\mathrm{trans}$, tidal heating becomes significant and our model surely underestimates the radius inflation during this phase.

We also note that the tidal dissipation constant $Q_\mathrm{p}$ is fixed to a fiducial value of $10^5$ for all simulations (Table~\ref{tab:fixed_param}). While a higher $Q_\mathrm{p}$ would assuredly delay the onset of tidal migration (shifting $t_\mathrm{trans}$ to later times), it would not extend the duration over which the simulated eccentricity matches the observed value. The eccentricity evolution would simply cross the observationally compatible region at a later time, maintaining a similar crossing duration. Crucially, this means that the total log-likelihood $\log p$ (Eq.~(\ref{eq:logP})) would not be enhanced by adopting a higher $Q_\mathrm{p}$; in other words, a slower migration has no intrinsic advantage in our framework as long as the crossing occurs within the allowed age range. Given the poor observational constraints on $Q_\mathrm{p}$ for Neptune-mass planets, we chose not to vary this parameter, focusing instead on the initial conditions that more directly control the evolutionary pathway.

\subsection{Impact of the initial mutual inclination}
\label{sect:incl}

\begin{table}
\caption{Results of the $\left(a_0, \, M_\mathrm{p,\,0}, \, \Lambda\right)$ retrieval as a function of $i_\mathrm{mut,\,0}$.}
\centering
\setlength\tabcolsep{0pt}
\begin{tabular*}{\linewidth}{@{\extracolsep{\fill}} cccc }
\hline
$i_\mathrm{mut,\,0}$ [$^\circ$] & $\log\Lambda$ & $a_0$ [AU] & $M_\mathrm{p,\,0}$ [$M_\oplus$] \\
\hline
\hline
60 & $-4.72^{+0.12}_{-0.79}$ & $0.31^{+0.02}_{-0.10}$ & $22.00^{+0.29}_{-0.08}$ \\
70 & $-4.57^{+0.11}_{-0.40}$ & $0.35^{+0.01}_{-0.14}$ & $21.98^{+0.39}_{-0.05}$ \\
75 &  $0.39^{+0.35}_{-0.19}$ & $0.26^{+0.05}_{-0.02}$ & $21.97^{+0.13}_{-0.05}$ \\
80 &  $1.24^{+0.11}_{-0.08}$ & $0.31^{+0.01}_{-0.01}$ & $22.02^{+0.21}_{-0.09}$ \\
85 &  $1.59^{+0.41}_{-0.17}$ & $0.33^{+0.05}_{-0.02}$ & $22.05^{+0.26}_{-0.13}$ \\
\hline
\end{tabular*}
\label{tab:retrieval}
\end{table}

Showing the results for a fiducial initial mutual inclination in this way (Sect.~\ref{sect:resfid}) allows us to scrutinize the influence of the other parameters in isolation and deepen our understanding of the ZLK mechanism. We now expand our analysis beyond this case to assess how the initial mutual inclination impacts our exploration. For each explored value of $i_\mathrm{mut,\,0}$, we conducted the same comprehensive semi-Bayesian retrieval process described previously (Sect.~\ref{sect:fw}), with each refined grid comprising approximately $100\,000$ simulations. All other simulation parameters (Table~\ref{tab:fixed_param}) and observational constraints (Table~\ref{tab:constraints}) remain unchanged.

The results of our $\left(a_0, \, M_\mathrm{p,\,0}, \, \Lambda\right)$ retrieval as a function of $i_\mathrm{mut,\,0}$ are shown in Table~\ref{tab:retrieval}. Globally, we observe the same behaviors and reach similar conclusions as in Sect.~\ref{sect:resfid}, particularly regarding the minimal impact of atmospheric evaporation, corroborated by the consistent derived values of $M_\mathrm{p,\,0}$ across all explored initial mutual inclinations. Notable differences emerged when comparing higher initial mutual inclinations ($80^\circ$ and $85^\circ$) to the fiducial case. For these higher values, the maximum eccentricity reached during ZLK cycles $e_\mathrm{max}$ increases \citep{Katz2011}, necessitating larger initial separations $a_0$ due to angular momentum conservation as discussed in Sect.~\ref{sect:resfid}. This in turn leads to higher derived values of $\Lambda$ to maintain a roughly constant ZLK timescale $\tau_\mathrm{ZLK} \propto \Lambda/a^{3/2}$, as illustrated in Table~\ref{tab:retrieval}. Such a behavior does not hold anymore for the two lowest inclinations ($60^\circ$ and $70^\circ$), which is due to the much more chaotic nature of their evolutions, where SRFs become comparable in magnitude to the ZLK perturbation. The clear two-fold behavior showcased by the three highest $i_\mathrm{mut,\,0}$ (e.g., Fig.~\ref{fig:gj436_maxlike}) is not seen anymore in the low-inclination regime. In any case, the retrieval seems to favor configurations with much shorter ZLK timescales for them, yet with quite large uncertainty intervals.

\begin{figure}
\includegraphics[width=\linewidth]{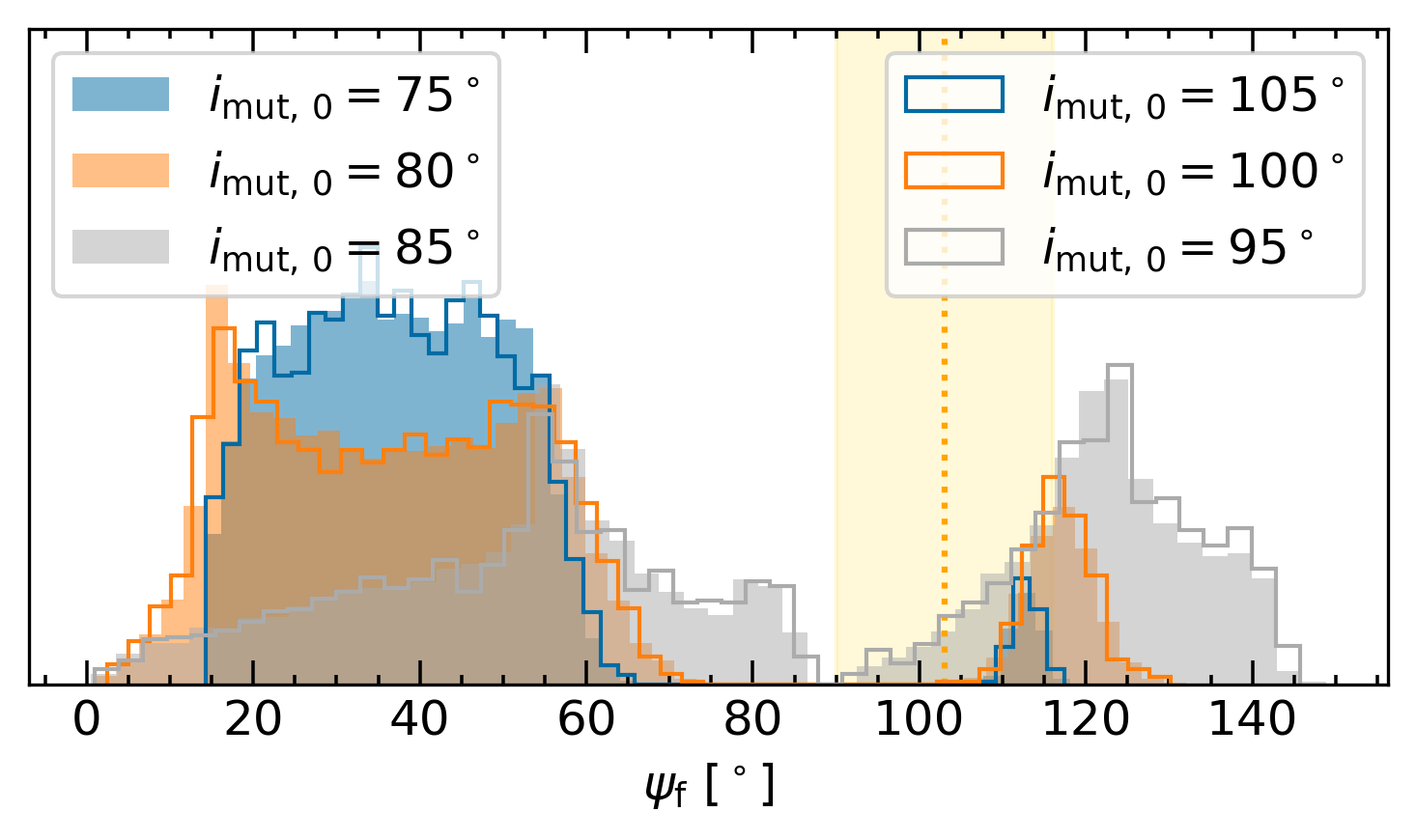}
\centering
\caption{Distribution of final spin--orbit angles as a function of different initial mutual inclinations, for simulations where a resonance was triggered and then exited. The orange shaded region is the observational constraint on $\psi$.}
\label{fig:gj436_psif_imut}
\end{figure}

The impact of initial mutual inclination is particularly striking when examining the distribution of final spin--orbit angles $\psi_\mathrm{f}$ after decoupling (Fig.~\ref{fig:gj436_psif_imut}). As $i_\mathrm{mut,\,0}$ increases and approaches $90^\circ$, the associated $\psi_\mathrm{f}$ distribution becomes wider, with a growing proportion of retrograde orbits, akin to the results of \citet{Fabrycky2007}. For the lowest investigated initial mutual inclinations ($60^\circ$ and $70^\circ$), no retrograde orbits emerged after decoupling from the perturber, which is why these distributions have been excluded from Fig.~\ref{fig:gj436_psif_imut}. Since these configurations cannot produce final spin--orbit angles compatible with the observed $\psi_\bullet$, they can be discarded as viable solutions, though we include their retrieval results in Table~\ref{tab:retrieval} for completeness.

To complement our analysis, we investigated whether mirror configurations with respect to $i_\mathrm{mut,\,0} = 90^\circ$ would yield similar results. The ZLK equations at the quadrupolar level exhibit symmetry with respect to $i_\mathrm{mut,\,0} = 90^\circ$, suggesting that a system with initial mutual inclination $i_\mathrm{mut,\,0}$ and another with $180^\circ - i_\mathrm{mut,\,0}$ should follow identical evolutionary paths for most orbital elements during the resonance phase \citep{Sidorenko2018,Lei2022}. We tested this theoretical expectation by selecting $5\,000$ random configurations for each prograde $i_\mathrm{mut,\,0}$ value ($60^\circ$, $70^\circ$, $75^\circ$, $80^\circ$, and $85^\circ$) that successfully triggered a ZLK resonance and subsequently decoupled within the simulation timeframe. We then simulated their mirror configurations with retrograde $i_\mathrm{mut,\,0}$ values ($120^\circ$, $110^\circ$, $105^\circ$, $100^\circ$, and $95^\circ$ respectively), totaling $25\,000$ additional simulations. Our numerical verification confirms the expected symmetry for most initial mutual inclinations during the resonance phase. The exception is at $i_\mathrm{mut,\,0} = 60^\circ$ (and its mirror at $120^\circ$), where we occasionally observe substantial differences---some systems would decouple while their mirrors remain locked in ZLK cycles throughout the simulation, for instance. This discrepancy results again from the more chaotic nature of lower-inclination configurations where SRFs become prevalent even during the coupled phase. 

Furthermore, while the symmetry holds during the resonance, it breaks after decoupling as the subsequent evolution becomes dominated by tidal effects, whose governing equations lack this symmetry. Despite this, our analysis reveals that the distributions of final spin--orbit angles for mirror configurations are remarkably similar to their prograde counterparts, as shown in Fig.~\ref{fig:gj436_psif_imut}. While individual mirror systems do not necessarily produce identical final spin--orbit angles (i.e., we observed some jitter between paired systems and even some cases where a system with $\psi_\mathrm{f} \simeq 60^\circ$ had a mirror counterpart with $\psi_\mathrm{f} \simeq 120^\circ$ and vice versa), these differences average out when examining the global distribution. Kolmogorov--Smirnov tests confirm this similarity with $p$-values of $> 0.5$, indicating no statistically significant difference between $\psi_\mathrm{f}$ distributions of mirror systems, for all pairs except $i_\mathrm{mut,\,0} = 60^\circ$ and $120^\circ$. Notably, configurations with $i_\mathrm{mut,\,0} = 120^\circ$ produced no retrograde final orbits, consistent with their prograde counterparts at $60^\circ$. Thus, they were also excluded from Fig.~\ref{fig:gj436_psif_imut}.

Based on our analysis of final spin--orbit angles (Fig.~\ref{fig:gj436_psif_imut}), we can now confidently identify the range of initial mutual inclinations that best satisfy the observational constraint on $\psi_\bullet$. Configurations with $i_\mathrm{mut,\,0} = 80^\circ$ and $85^\circ$ (and by extension, their retrograde counterparts at $100^\circ$ and $95^\circ$) produce substantial numbers of systems with final spin--orbit angles compatible with the observed value. We therefore conclude that initial mutual inclinations in the range $80^\circ - 100^\circ$ most successfully reproduce GJ~436~b's observed polar orbit while maintaining consistency with all other observational constraints. We did not directly explore initial mutual inclinations beyond $85^\circ$ (or below $95^\circ$ for retrograde configurations) because we observed increasing discrepancies between \J{} simulations and $N$-body codes \citep{Trani2023} near $i_\mathrm{mut,\,0} \simeq 90^\circ$. Nevertheless, the clear trend of increasing retrograde final orbits as $i_\mathrm{mut,\,0}$ approaches $90^\circ$ from either direction strongly suggests that this central value would also be compatible with observations.

\section{Plausible location of GJ~436~c}
\label{sect:detect} 

\subsection{Exploring the detection limits}
\label{sect:detlim}

It is finally worth putting our results in an observational context to pinpoint the possible location of the concealed companion GJ~436~c. Given the decades of RV data acquired for the system, a certain fraction of the parameter space can already be excluded. To perform such an analysis, we retrieved all publicly available HARPS \citep{Mayor2003} data collected between 2003 and 2020 from the ESO archive\footnote{\url{https://archive.eso.org/wdb/wdb/adp/phase3_spectral/form}}, encompassing three distinct observational programs: HARPS GTO (Mayor, 072.C-0488), ``Transit of telluric exoplanets orbiting M dwarfs'' (Bonfils, 082.C-0718), and ``The HARPS search for planets around M dwarfs'' (Bonfils, 1102.C-0339). The RVs were extracted using a template-matching code (\texttt{NAIRA}), specifically optimized for M dwarfs, previously applied in multiple HARPS-based exoplanet searches \citep[e.g.,][]{AstudilloDefru2015}. The RVs were then binned per night and corrected from secular acceleration. Our final set consists of 171 nights from HARPS03, spanning $1\,532$ days, and 18 nights from HARPS15, spread over/covering only 41 nights. This results in a total observational baseline of $5\,157$ days, with a $3\,584$-day gap. The offset between the two datasets is determined through the modeling of GJ~436~b. Since both datasets are extracted from different templates, the offset is statistically consistent with zero.

After removing the RV contribution of GJ~436~b \citep[using the planetary parameters from][]{Bourrier2018}, we compute detection limits based on the Generalized Lomb--Scargle method proposed by \citet{Zechmeister2009}. This involves an injection--recovery test of a sinusoidal signal in the RVs at 12 equispaced phases, for each tested period. For this study, we adopt the most conservative detection limit \citep[see][for details]{Mignon2024}, defined as the largest injected amplitude required for the periodogram signal to reach the 99\% significance level at the period of the injected signal. 

We also mention the collection of supplementary observational detection limits stemming from the direct images of NACO \citep{Brandner2002} from the ESO archive (Apa\"{i}, 081.C-0430). However, they were discarded because less stringent than the RV constraints.

\subsection{Constraining the parameter space}

\begin{figure}
\includegraphics[width=\linewidth]{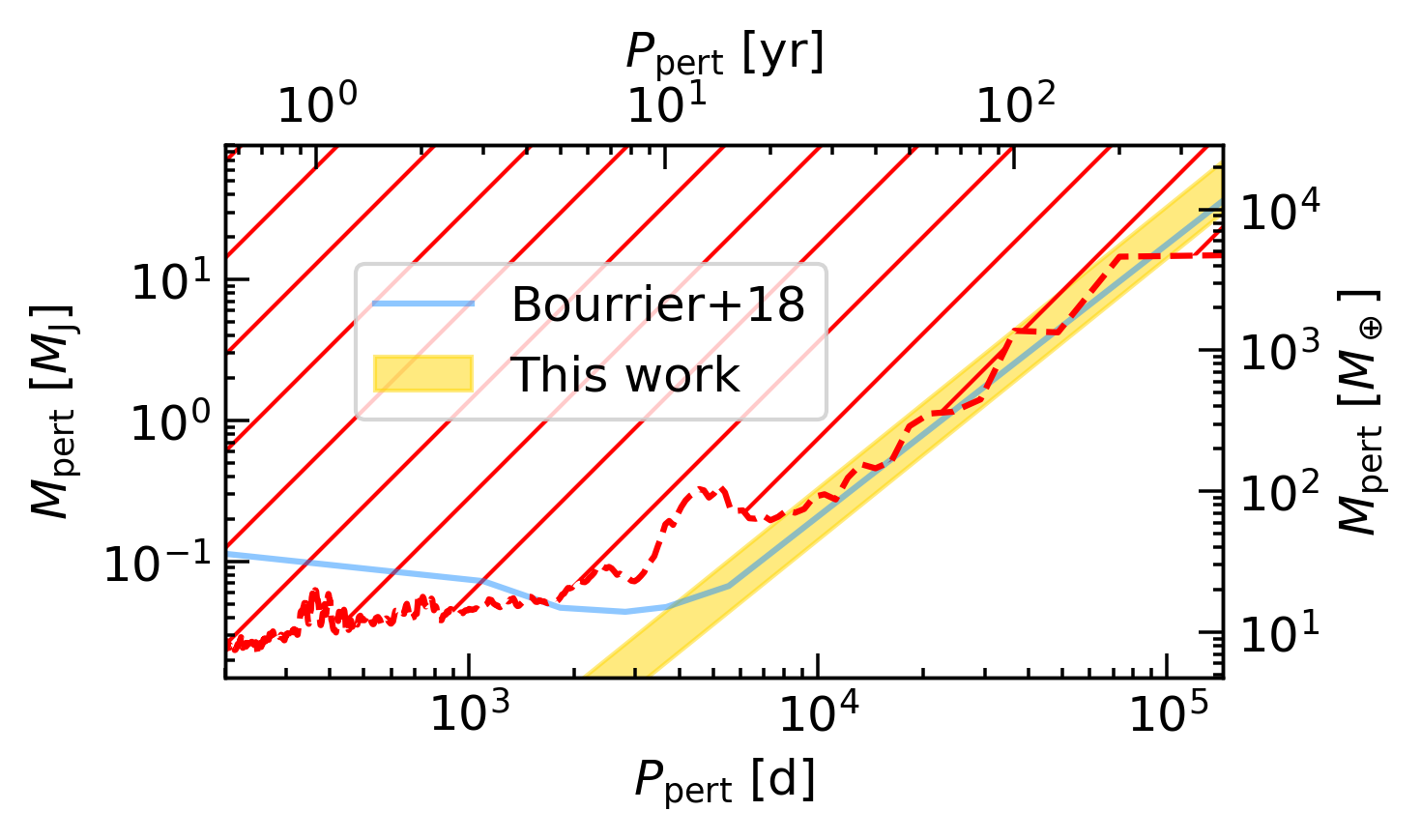}
\centering
\caption{Possible parameter space where GJ~436~c could hide in the mass--period plane, assuming a circular orbit. The orange shaded streak is the compatible configuration from the derived values of $\Lambda$ for $i_\mathrm{mut,\,0}$ values of $80^\circ$ and $85^\circ$. The blue line is the result from \citet{Bourrier2018}. The red hatched region, delimited by the red dashed line, is excluded by RV measurements.}
\label{fig:gj436_detection}
\end{figure}

Figure~\ref{fig:gj436_detection} shows the possible values for the properties of planet c from our exploration. The red hatched region corresponds to the detection limits excluded from archival RV measurements (Sect.~\ref{sect:detlim}). The bright streak includes the retrieved values of $\Lambda$ for $i_\mathrm{mut,\,0}$ values of $80^\circ$ and $85^\circ$ (Table~\ref{tab:retrieval}), as they are the only explored initial mutual inclinations satisfying the observational constraint on the spin--orbit angle (along with their mirror counterparts $100^\circ$ and $95^\circ$, which would yield the same retrieved values of $\Lambda$ thanks to the symmetry, see discussion in Sect.~\ref{sect:incl}). The two compatible values of $\Lambda$ were merged, assuming their dependence to $i_\mathrm{mut,\,0}$ is continuous for this high-inclination regime, which seems justified in light of Table~\ref{tab:retrieval}. Left of the bright streak and for a given $a_0$, the decoupling occurs too quickly: the orbit is already circularized after $t_{\bullet\downarrow}$. To the right of it, we can see the resonance is either too slow (no decoupling after $t_{\bullet\uparrow}$) or does not even occur at all, in line with the findings of \citet{Beust2012}. 

Our compatible streak is consistent with the initial proposal of \citet{Bourrier2018} for long periods. The shape of their solution departs from ours for the lowest periods, where our compatible streak diverges from their results. This divergence occurs because these configurations correspond to low ratios of outer to inner orbital angular momentum (of order a few), a regime where the secular approximation begins to break down \citep{Mangipudi2022} and higher-order effects become significant. While our quadrupole-only treatment cannot capture such contributions, \citet{Beust2012} showed that the corrected compatible parameter space in this regime would follow $M_\mathrm{pert} \propto P_\mathrm{pert}^{-1/3}$ (locus of constant angular momentum). Crucially, this correction would push the viable solutions at $P_\mathrm{pert} < 10\,\mathrm{yr}$ into the region already excluded by our RV detection limits (as seen in Fig.~\ref{fig:gj436_detection}). Therefore, while our quadrupolar approximation does introduce differences compared to \citet{Bourrier2018} at short periods, these differences minimally affect our conclusions about the viable parameter space for GJ~436~c. It is also worth mentioning that our solution is wider compared to \citet{Bourrier2018}, which is due to our semi-Bayesian exploration framework yielding uncertainty margins (while they find compatible configurations by trial and error on a coarse grid). Plus, we virtually include the range of $i_\mathrm{mut,\,0}$ between $80^\circ$ and $85^\circ$ (while they fix $i_\mathrm{mut,\,0} = 85^\circ$). In reality, our compatible streak should be even larger so as to encompass systems with $i_\mathrm{mut,\,0}$ up to $90^\circ$, which are out of our numerical reach. 

Importantly, our compatible parameter space does intersect with the refined RV detection limits at $P_\mathrm{pert} > 200 - 300\,\mathrm{yr}$, from which we effectively rule out stellar companions, as they would fall entirely within the excluded region. Even brown dwarf companions appear highly unlikely, as they would require fine-tuned combinations of mass and orbital parameters to remain consistent with both the ZLK constraints and the nondetection in RV data. The viable parameter space thus strongly favors (cold) planetary-mass companions. In any case, there is still a window of opportunity to eventually detect this rogue companion, possibly by future missions, or JWST \citep{Gardner2006} via direct imaging. Figure~\ref{fig:gj436_detection} could serve for observers as a reliable basis for subsequent planet hunting campaigns.

\section{Discussion and conclusion}
\label{sect:conclu}

The analysis described in this work comes with a certain number of caveats. The main limitation would be the first-order formalism we employed for tidal forces, which could be further refined to include frequency-dependent inertial waves flowing from the rheological properties of the planet \citep{Andre2017,Andre2019,Dhouib2024}. In relation to this, while our calculation of the intrinsic temperature, $T_\mathrm{int}$, includes contributions from atmospheric and core cooling/contraction as well as radiogenic heating from heavy element decay \citep[following tabulated models from][]{Mordasini2020}, it currently neglects the heat generated by tidal friction. The planet's internal energy budget in \J{} thus accounts for three of the main contributions (cooling, radioactive decay, and irradiation), but missing the tidal component\footnote{Ohmic dissipation should also be taken into account, especially in the context of mini-Neptunes \citep{Pu2017}.}. This omission means that $T_\mathrm{int}$ does not spike during the rapid migration phase (as might be expected in Fig.~\ref{fig:gj436_maxlike}), potentially underestimating planet inflation at $t_\mathrm{trans}$. Additionally, the planet structure handling could have been better than imposing default values for most of the parameters (Sect.~\ref{sect:bulk}), but the exploration tells us that erosion is notably ineffective. A more elaborate prescription would have just resulted in a systematic shift in the radius for all configurations, perhaps slightly changing the inferred values because of the feedback of tides. Finally, our treatment of the stellar bulk parameters is quite crude. Stars experience mass loss, inflation, and contraction along their main sequences, which should be taken into account given the sensitivity of tidal dissipation to such effects. In particular, a more realistic evolution of the spin--orbit angle is anticipated, should these processes be implemented.

In this work, we applied the novel semi-Bayesian exploration framework of \J{} to GJ~436~b. It allowed us to constrain its formation location around $0.3\,\mathrm{AU}$ and pinpoint the properties of a potential distant companion, identified to be planet-mass, which could explain this enigmatic system's observed characteristics on the basis of a late-stage HEM scenario. Our exploration reveals that GJ~436~b likely formed far enough from its host star to avoid substantial atmospheric erosion during the star's highly active youth. The planet's current eccentric and polar orbit strongly supports a scenario where a hidden companion induced ZLK oscillations with initial mutual inclinations of $80^\circ - 100^\circ$, eventually driving GJ~436~b to its present-day location. The minimal impact of atmospheric mass loss throughout this history demonstrates that atmospheric evolution and orbital dynamics must be treated simultaneously to accurately reconstruct the evolutionary pathways of close-in planets.

We draw attention to the fact that our framework accounts for the statistical likelihood of observing GJ~436~b in its current state. We acknowledge that our proposed late migration scenario results in the planet matching its observed eccentricity for less than 1\% of the system's lifetime (as seen in Fig.~\ref{fig:gj436_maxlike}). However, this apparent fine-tuning is properly considered through our temporal integration of individual probabilities, expressed via Eq.~(\ref{eq:logP}), and also elaborated on in Appendix~\ref{app:bayes}). It naturally favors evolutionary scenarios that maintain compatibility with observations over extended periods, while penalizing those that only briefly match the current system parameters. Importantly, we find this late migration scenario substantially more conceivable than the alternative ``no-migration'' hypothesis. The latter would require an unrealistically high tidal dissipation factor \citep[$Q_\mathrm{p} \gtrsim 10^6$,][]{Mardling2008} to maintain the observed eccentricity over Gyrs, offers no explanation for the polar spin--orbit angle, and would necessitate an implausibly massive initial planet to survive the intense evaporation during the star's energetic infancy. Forming such massive planets at $\sim\,0.03\,\mathrm{AU}$ around M dwarfs appears to be disfavored by both theory and observations \citep{Laughlin2004,Endl2006,Burn2021,Schlecker2022}. In contrast, our late migration scenario explains all observed properties, albeit requiring an unseen companion, and set in a configuration that is consistent with current detection limits (Sect.~\ref{sect:detect}).

Nonetheless, the inferred initial configuration of GJ~436~b at $\sim\,0.3\,\mathrm{AU}$ with a companion on a near-perpendicular orbit raises important questions about the system's primordial architecture. Such a hierarchical two-planet system with high mutual inclination is certainly not a typical outcome of standard planet formation scenarios, possibly representing the weakest aspect of our hypothesis from a physical standpoint. While we do not directly address the origin of this primordial mutual inclination in this work, several mechanisms in the literature could potentially explain such a configuration. One possibility involves formation in a warped protoplanetary disk with misaligned inner and outer components. In this scenario, a massive planet carving a gap could break the disk into two misaligned sections \citep{Nealon2018,Nealon2019,Zhu2019}. Nevertheless, this mechanism preferentially tilts the inner disk, conflicting with our assumption of an initially aligned inner orbit, and would require an additional planet not considered in our analysis. Alternatively, planet--planet scattering could produce the required architecture if multiple cold giant planets underwent disruptive dynamical encounters, leaving a survivor on a highly inclined orbit \citep{Rasio1996,Chatterjee2008,Beauge2012,Petrovich2014}. However, the ability to generate near-perpendicular configurations through such events remains uncertain. 

Stellar flybys during the early cluster phase could also explain such a configuration by violently tilting the outer orbit \citep{Malmberg2011}, although this would need to occur shortly after disk dispersal to match our inferred initial conditions, as the disk's presence strongly dampens external perturbations \citep{Picogna2014}. Flybys could directly deposit planets on wide orbits as well, following episodes of enhanced scattering \citep{Bailey2019}, as demonstrated comprehensively by \citet{Izidoro2025} via instabilities that end up leveraging the dense birth environment. Such far-orbiting companions would be dynamically decoupled from inner planets, naturally allowing for substantial mutual inclinations. However, most studies of these cluster-based mechanisms have focused on Sun-like stars and infer typical orbital separations $> 100\,\mathrm{AU}$. The relevant distances for cluster dynamics scale proportionally as $M_\star^{1/3}$ (Hill sphere), potentially bringing the effective range down to $> 70\,\mathrm{AU}$ and closer to our inferred companion orbital distances. Whether these mechanisms can operate efficiently at the closer end of our viable parameter space (a few tens of AU) remains an open question. Encouragingly, a growing number of exoplanet systems exhibit similar architectures, with close-in small planets being highly misaligned in relation to distant giants, including HAT-P-11 \citep{Winn2010,Yee2018}, $\pi$~Men \citep{Jones2002,Xuan2020,DeRosa2020}, and HD~73344 \citep{Sulis2024,Zhang2025}. These observations confirm that our inferred configuration, while appearing exotic at first glance, could indeed be possible in nature. The challenge now is understanding the potentially multistage processes that produce such systems.

This study has important implications for understanding the broader population of exoplanets at the edge of the hot Neptune desert. The eccentric, misaligned systems forming the ``Neptunian ridge'' described by \citet{CastroGonzalez2024} might share similar evolutionary histories involving late migration, explaining both their survival and their unusual orbital configurations. GJ~436~b could therefore be a prototype of a distinct population of planets whose histories are shaped by the complex interplay between atmospheric processes and dynamical perturbations. Future works will extend this analysis to a larger sample of Neptune-sized planets to test whether HEM is a common pathway for planets populating the boundaries of the desert (Bourrier et al., under review). While our current constraints on GJ~436~c provide a roadmap for future observational campaigns, direct detection would provide the ultimate validation of this scenario.

\begin{acknowledgements}
We are grateful to the anonymous referee for their review, which improved the quality of our manuscript. We offer our thanks to Elsa Bersier (ESBDI, Geneva) for designing the \J{} logotype. We thank Aline Vidotto for insightful conversations. The authors made use of the Claude AI assistant (Anthropic, 2024) for code optimization. This work has been carried out within the framework of the NCCR PlanetS supported by the Swiss National Science Foundation under grants 51NF40$\_$182901 and 51NF40$\_$205606. This project has received funding from the European Research Council (ERC) under the European Union's Horizon 2020 research and innovation programme (project {\sc Spice Dune}, grant agreement No 947634; grant agreement No 730890). This material reflects only the authors' views and the Commission is not liable for any use that may be made of the information contained therein.
\end{acknowledgements}

\bibliographystyle{aa} 
\bibliography{biblio}

\begin{appendix}

\section{\J{} 101: Use cases}
\label{app:jade_101}

\begin{figure}[h]
\includegraphics[width=\linewidth]{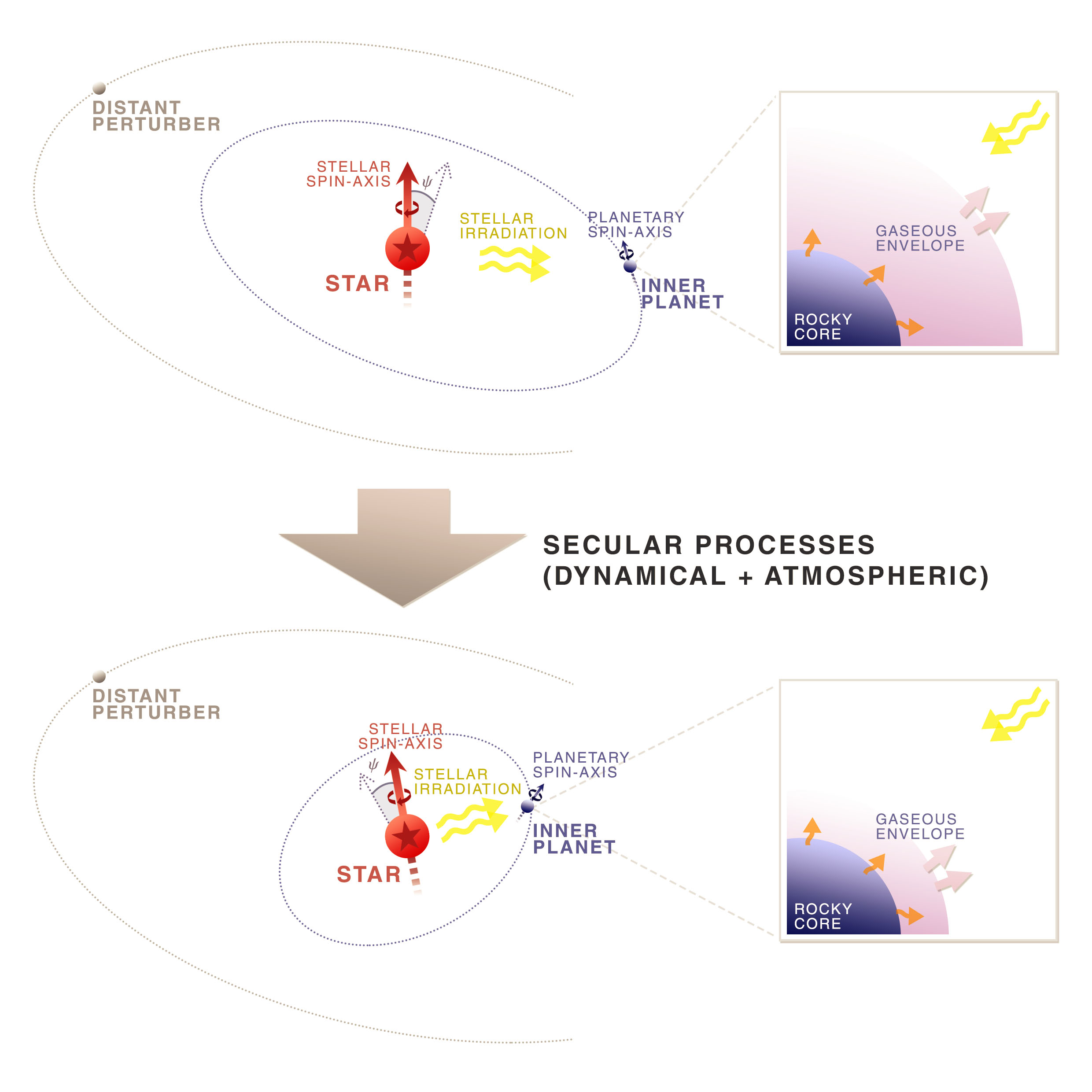}
\centering
\caption{Schematic representation of the \J{} code's key components and processes. The system shows a central star (red) with two orbiting bodies: an inner planet (blue) with a gaseous envelope and a solid core, and a distant perturber (gray). The sketch illustrates the evolution of the system across two time steps, demonstrating secular changes in the inner planet's orbital parameters and spin-axis orientation, while the outer orbit remains fixed. The zoomed insets depict the inner planet's atmospheric structure, showing core heating (orange) on one hand, and atmospheric escape (purple) processes under stellar irradiation (yellow) on the other hand. The three-dimensional (3D) spin-orbit angle, $\psi$, is highlighted in light blue. See \citet{Attia2021} for more details.}
\label{fig:jade_global}
\end{figure}

\begin{figure}
\includegraphics[width=\linewidth]{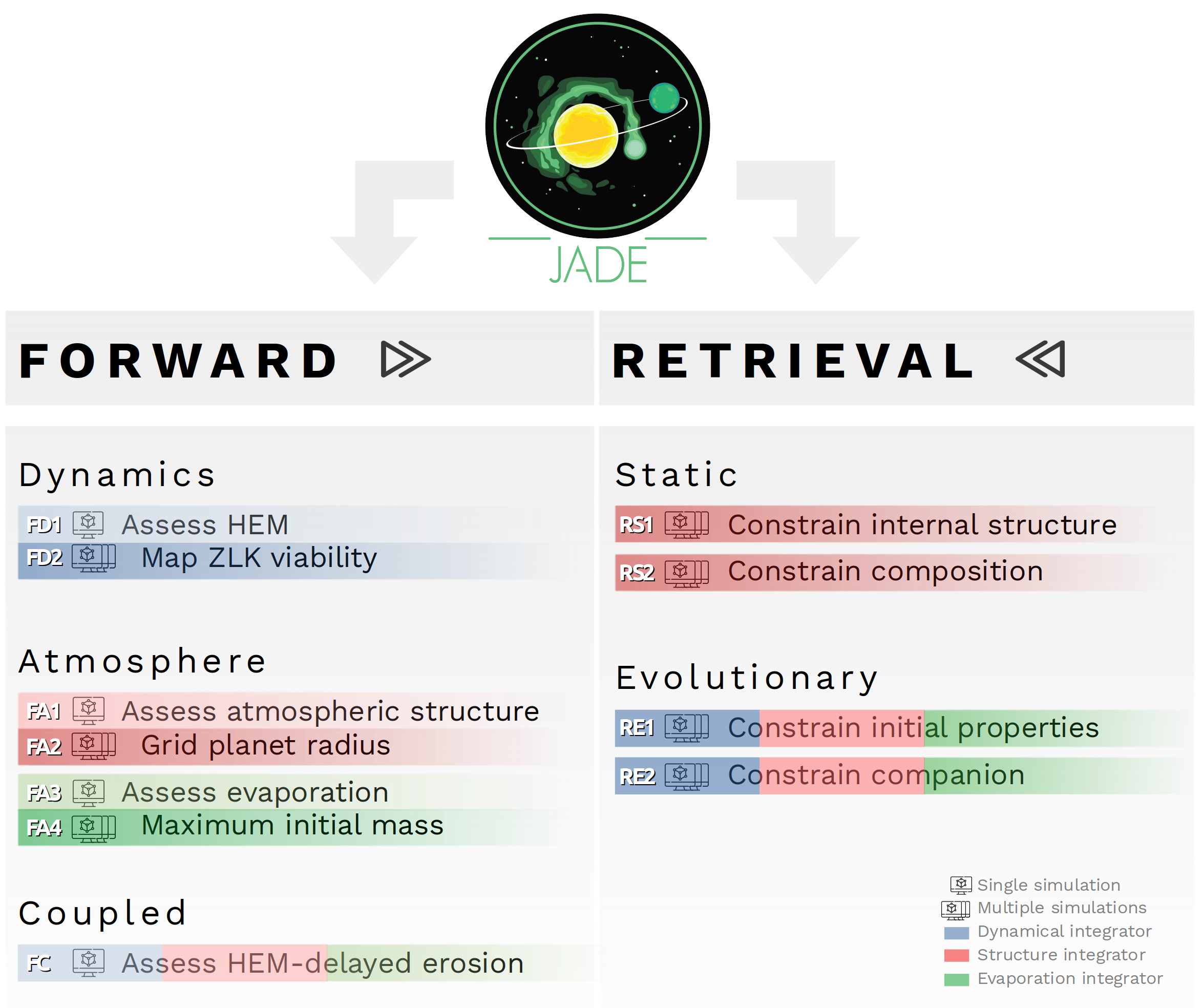}
\centering
\caption{Overview of the different use cases for the \J{} code, structured into forward and retrieval modeling. For every use case, an identification code is specified on its very left, and an icon shows whether it requires a single simulation or multiple ones. The shaded rectangles behind the use cases denote which integrators are involved.}
\label{fig:jade_use_cases}
\end{figure}

The typical configuration simulated by \J{} is depicted in Fig.~\ref{fig:jade_global}. One of the major strengths of the \J{} code is its versatility, capable of tackling a wide spectrum of questions. Figure~\ref{fig:jade_use_cases} illustrates the variety of use cases (FD1, FD2, FA1...) that are achievable with it, covered by pedagogical tutorials within the first distribution of the software linked in the first page of this article.

The \J{} code was originally designed to identify families of secular behaviors and pinpoint the influential mechanisms at play in the framework of a coupled dynamics--atmosphere treatment. The FD1, FA1, FA3 use cases, and essentially their combination FC (Fig.~\ref{fig:jade_use_cases}, left panel) embody this qualitative philosophy. They represent the foundational blocks on which the other use cases are built. Starting from a specified set of orbital and bulk parameters, \J{} can forwardly simulate the system's dynamical evolution including all of the relevant contributions for ZLK-induced HEM (FD1), as well as the close-in planet's evaporative evolution (FA3), characterize its atmospheric structure at a given time (FA1) by computing various profiles (e.g., temperature--pressure, mass distribution), or amalgamate the above for a fully coupled, realistic evolutionary history (FC). 

Running sequentially several simulations pertaining to one of these use cases allows us to consider additional, more quantitative application scenarios. For example, by iterating FD1 over some system parameters in a grid-like structure, we can locate the regions of the parameter space where the ZLK resonance can surpass the action of SRFs and be effectively triggered, which forms the FD2 use case \citep[an implementation is described in][]{Hagelberg2023}. In practice, such ``ZLK viability'' maps are generated using analytical expressions and later validated with \J{} simulations coarsely sampling the parameter space, for efficiency purposes. On the atmospheric side, we can perform multiple simulations where evaporation is the exclusive operating process (FA3) launched on a range of different initial planet masses, assuming a fixed orbit for all of them; namely, the one corresponding to the present-day configuration. In doing so, the initial planet mass yielding a mass compatible with the current one at the age of the system can be interpreted as the maximum initial mass (FA4). Indeed, an unchanged close-in orbit during the entire planet's lifetime (in situ formation or early-on disk migration) represents the locus of maximum possible erosion, and any initial mass that is too high to evaporate enough to match that of the present-day can definitely be excluded\footnote{Technically, this interpretation is correct only if the migration occurs inward. In practice, this condition is always satisfied for our close-in systems.} (an application will be soon available in Livingston et al.~\textit{in prep.}). Finally, individual snapshots of what the planet structure looks like (FA1) for different sets of bulk and irradiation parameters can be used to construct a grid of atmospheric models (FA2, Sect.~\ref{sect:grid}), which can prove helpful in deriving accurate mass--radius relationships (Sect.~\ref{sect:mr}) and capitalizing on all the benefits they bring.

Recently, \J{} has been also employed in retrieval mode, meaning that today's system characteristics allow us access to all sorts of hitherto unknown properties (Fig.~\ref{fig:jade_use_cases}, right panel). Indeed, the \J{} code is able to constrain a planet's internal structure (iron core, silicate mantle, and volatile atmosphere mass fractions, RS1) as well as the composition of its H/He-dominated atmosphere (hydrogen $X$, helium $Y$, and metal $Z$ mass fractions, RS2) with the observational mass and radius in a Markov chain Monte Carlo (MCMC) approach. This ``static'' (present-day) bulk characterization often serves as a first step prior to the ``evolutionary'' parameter space exploration, the latter ultimately enabling us to constrain the close-in planet's birthplace (RE1) and outer companion (RE2) to reveal credible pathways compatible with the currently measured parameters. The semi-Bayesian infrastructure in which RE1 and RE2 are embedded is detailed in Sect.~\ref{sect:fw}. Such an exploration is made possible among others through the upstream reduction of the parameter space to inspect via FD2 and FA4 by ruling out, from the outset, all configurations that quench ZLK cycles on the one hand, and that are too massive to sufficiently erode on the other hand. Critically, even the most limited of the parameter spaces would require a considerable number of simulations for a meaningful exploration due to the chaotic nature of three-body dynamics and, what is more, the nonlinear intervention of atmospheric feedback. This is precisely where FA2 intervenes so as to dramatically decrease the computation time: the atmospheric properties, in particular the self-consistently derived planet radius, are interpolated from a precomputed grid, bypassing the usage of the structure integrator, which is by far the biggest computational bottleneck of \J{}. This way, a typical fully coupled run can be made in around ten minutes, authorizing large explorations on computer clusters for instance. The quantitative performance tests of such an approach are presented in Appendix~\ref{app:perf}, and its results on our flagship system, GJ~436, are shown in Sect.~\ref{sect:explo}.

\section{Performance of the mass--radius relationships}
\label{app:perf}

\begin{figure}[h]
\includegraphics[width=\linewidth]{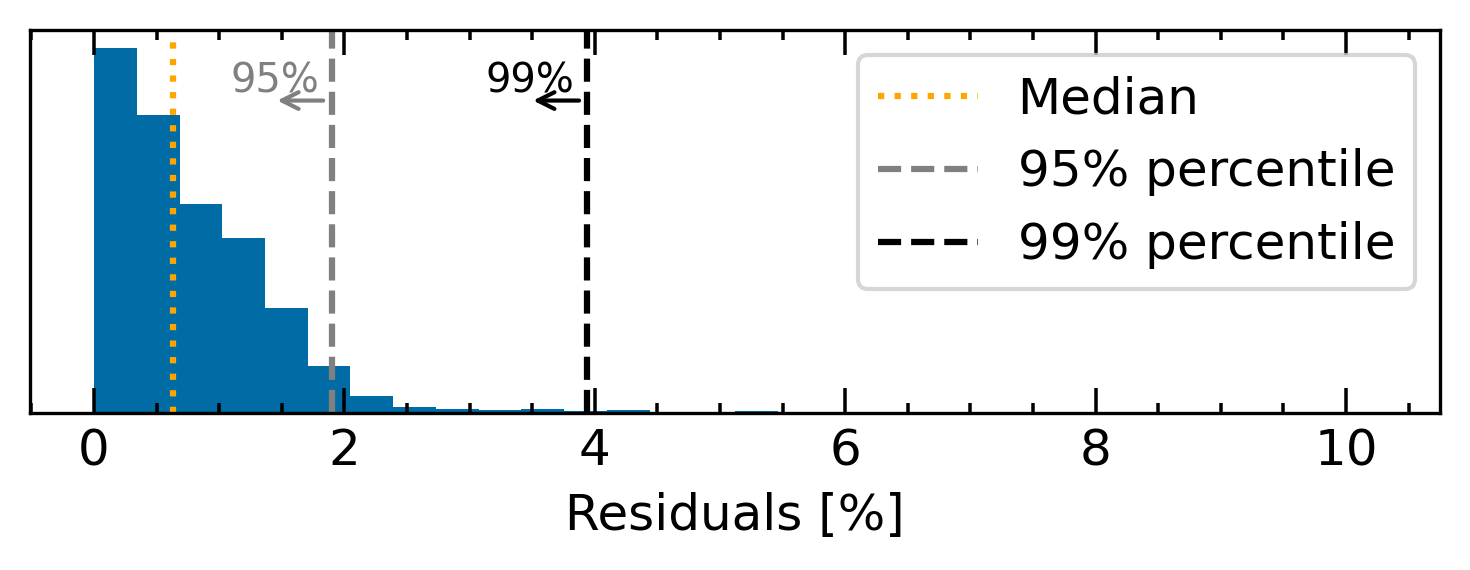}
\includegraphics[width=\linewidth]{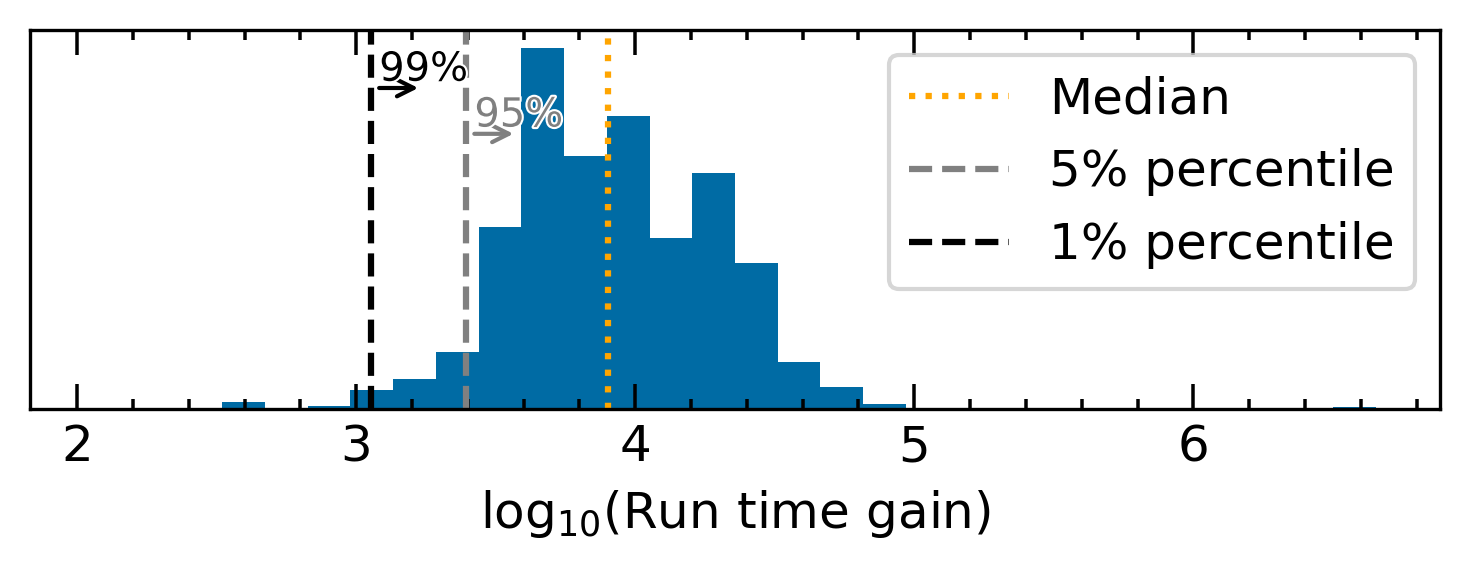}
\centering
\caption{Performance assessment of the derived mass--radius relationship for GJ~436~b.
\textbf{Top:} histogram of residuals. Are also shown the median, the 95\%, and 99\% quantiles.
\textbf{Bottom:} histogram of run time gains. Are also shown the median, the 5\%, and 1\% quantiles.}
\label{fig:gj436_mr_perf}
\end{figure}

Even though the analytical radii of Sect.~\ref{sect:mr} seem to be consistent with the ones directly generated by \J{} upon visual inspection (Fig.~\ref{fig:gj436_mr}), it is necessary to quantify their deviation to assess the performance of our mass--radius relationships. To this effect, a control sample of $10\,000$ $\left(M_\mathrm{p}, \, T_\mathrm{eq}, \, T_\mathrm{int}\right)$ 3D-points are randomly drawn from the same domains of definition set for the internal structure grid (Sect.~\ref{sect:grid}). For each point, the analytical radius $R_\mathrm{p, \, ana}$ is calculated from Eq.~(\ref{eq:MR}) and then compared to the ``true'' radius $R_\mathrm{p, \, mod}$, modeled by the structure integrator of the \J{} code, from which a residual metric is derived using the unsigned relative error between them

\begin{equation}
\mathrm{Residual} = \left\lvert \frac{R_\mathrm{p, \, mod} - R_\mathrm{p, \, ana}}{R_\mathrm{p, \, mod}} \right\rvert.
\end{equation}

\noindent Figure~\ref{fig:gj436_mr_perf} (top panel) shows the histogram of residuals for the $10\,000$ random points, manifesting an excellent accuracy. The median residual is 0.6\%, which is very satisfying compared to the typical residuals such mass--radius relationships yield (usually considered satisfactory if lower than 5\%). 95\% (respectively 99\%) of the control sample has a better residual than 1.9\% (respectively 3.9\%), meaning that the error resulting in using Eq.~(\ref{eq:MR}) instead of the integrator is always safely negligible. Further enhancement, the run time gain when computing $R_\mathrm{p, \, ana}$ versus $R_\mathrm{p, \, mod}$ is utterly dramatic (Fig.~\ref{fig:gj436_mr_perf}, bottom panel). In 99\% of cases, the former is faster than the latter by a factor of $10^{3.1} \simeq 1\,300$, and on average, by a factor of $10^{3.9} \simeq 8\,000$.

\section{Posterior distribution of the constrained parameters}
\label{app:bayes}

We use $\theta^\mathrm{I}_t$ to denote the orbital parameters of the inner parameters at the time of the observations and $y$ represents the data. The posterior distribution is 

\begin{equation}
    p(\theta^\mathrm{I}_t \mid y) = \frac{p(y \mid \theta^\mathrm{I}_t ) p (\theta^\mathrm{I}_t)}{p(y)}, 
\label{eq:posterior}
\end{equation}

\noindent where $p(\theta^\mathrm{I}_t)$ is the prior distribution on the orbital elements. Eq.~(\ref{eq:posterior}) gives the classical posterior distribution of the orbital elements fitted onto the data. In principle, the same could be done to obtain the posterior distribution of the initial position of the inner and outer planets ($\theta^\mathrm{I}$, $\theta^\mathrm{O}$), as well as the time from the initial starting point $t$

\begin{equation}
    p(\theta^\mathrm{I}, \theta^\mathrm{O},t \mid y) = \frac{p(y \mid \theta^\mathrm{I}, \theta^\mathrm{O},t ) p (\theta^\mathrm{I}, \theta^\mathrm{O},t)}{p(y)}. 
\end{equation}

\noindent Indeed, for each $\theta^\mathrm{I}$, $\theta^\mathrm{O}$, we could propagate the equation in time to get the estimated current properties of the inner planet $\Phi(\theta^\mathrm{I}, \theta^\mathrm{O},t)$, setting $\theta^\mathrm{I}_t$ in Eq.~(\ref{eq:posterior}) as this propagated value. 

This approach is approximated here by considering $\theta^\mathrm{I}_t$ as the data, which follows a Gaussian distribution. Denoting by $n$ the number of components of $\theta^\mathrm{I}_t$, denoting the uncertainty on each component $k$ by $\sigma_k$ and assuming that there is no correlation between the parameter components, the likelihood then becomes 
\begin{equation}
    p(\theta^\mathrm{I}_t \mid\theta^\mathrm{I}, \theta^\mathrm{O},t ) = \frac{1}{\sqrt{2\pi}\prod_{k=1}^n \sigma_k} \exp\left(-\frac{1}{2} \sum_{k=1}^n \frac{\left(\theta^\mathrm{I}_{t,k} -\Phi(\theta^\mathrm{I}, \theta^\mathrm{O},t )_k \right)^2 }{\sigma_k^2} \right).
\label{eq:likelihood_gauss}
\end{equation}

\noindent The likelihood function employed in this work (Eq.~(\ref{eq:like})) is none other than Eq.~(\ref{eq:likelihood_gauss}) marginalized over time.

We further add a prior on $\theta^\mathrm{I}$, $\theta^\mathrm{O}$, $t$ (uninformative in this study), so that their posterior distribution is approximated by 

\begin{equation}
    p(\theta^\mathrm{I}, \theta^\mathrm{O},t \mid \theta_I^t) = \frac{p(\theta^\mathrm{I}_t \mid\theta^\mathrm{I}, \theta^\mathrm{O},t ) p (\theta^\mathrm{I}, \theta^\mathrm{O},t)}{p(y)}. 
\label{eq:posterior_approx}
\end{equation}

\noindent Integrated over time, Eq.~(\ref{eq:posterior_approx}) is used to constrain the different quantities described in Sect.~\ref{sect:explo}.

\section{Additional figures}

\begin{figure*}[!b]
\includegraphics[width=0.83\linewidth]{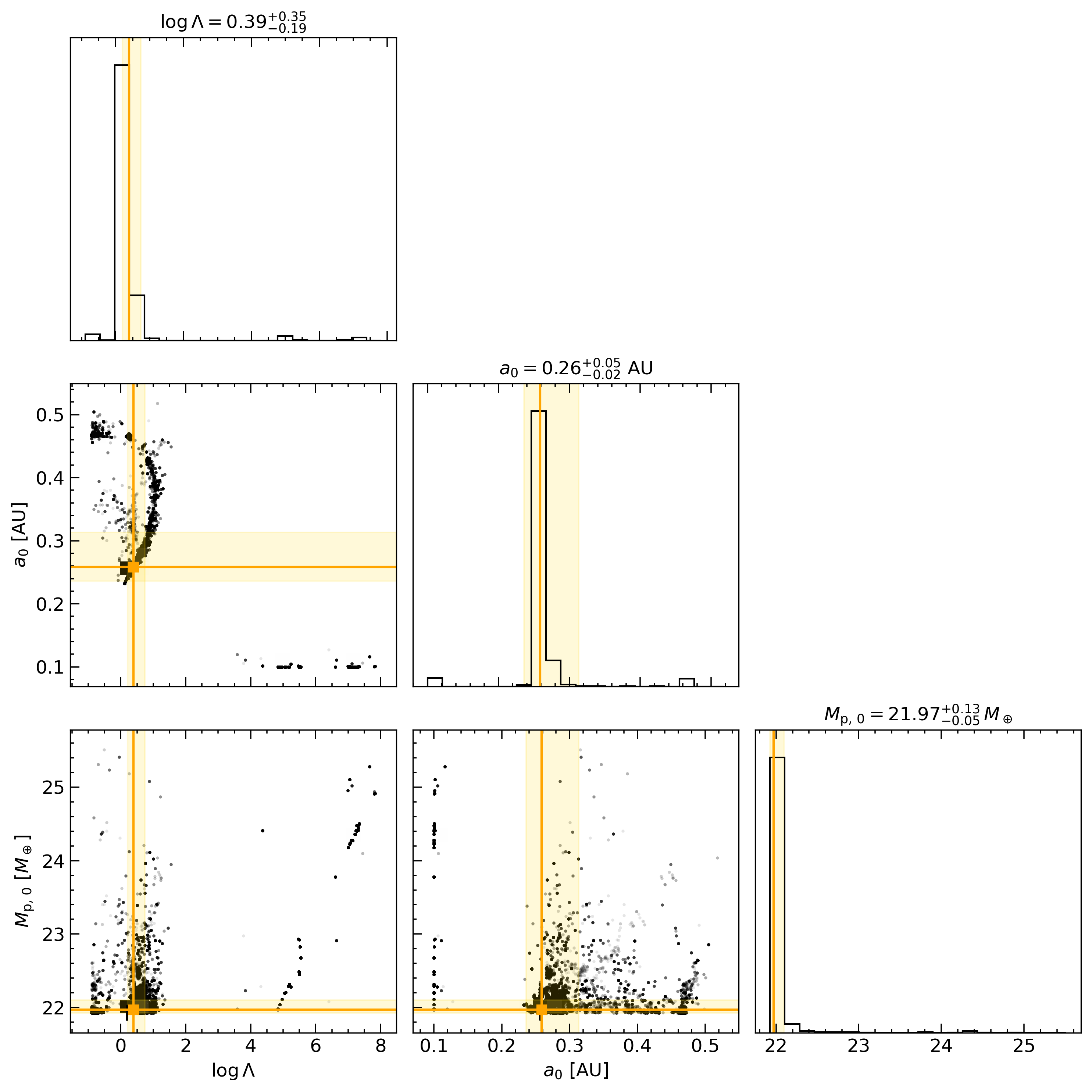}
\centering
\caption{Corner plot for the jump parameters $\left(a_0, \, M_\mathrm{p,\,0}, \, \Lambda\right)$ of the fiducial exploration ($i_\mathrm{mut,\,0} = 75^\circ$, Sect.~\ref{sect:resfid}). Orange lines indicate the medians of the posterior distributions, and the orange shaded regions their 68\% highest-density intervals (HDI). Their numerical values are reported on top of the 1D histograms. Here, $\log$ denotes the natural logarithm.}
\label{fig:gj436_corner}
\end{figure*}

\begin{figure*}[!b]
\includegraphics[width=0.83\linewidth]{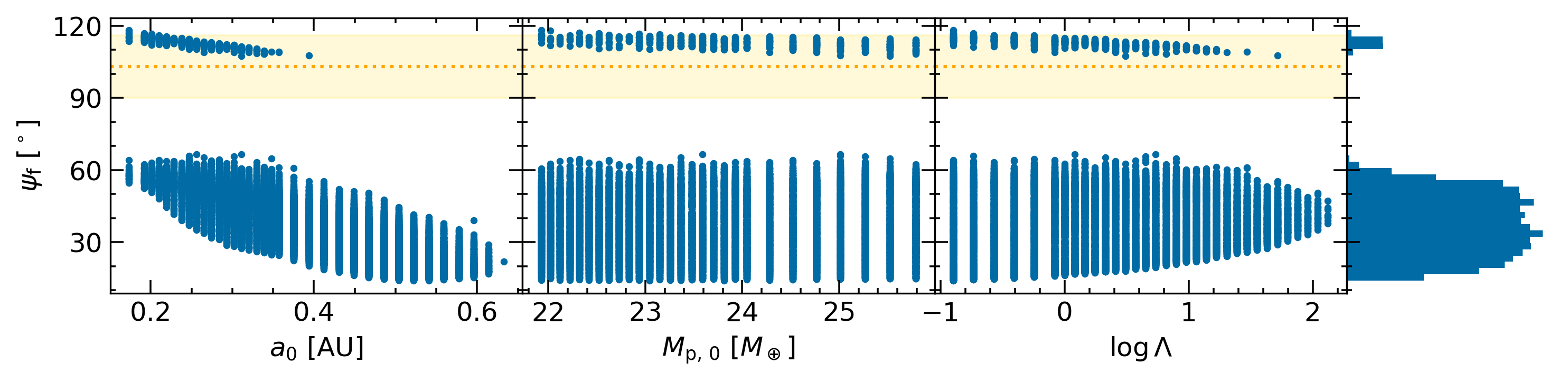}
\centering
\caption{Distribution of final spin--orbit angles as a function of the jump parameters, for the simulations where a resonance was triggered and then exited in the fiducial exploration ($i_\mathrm{mut,\,0} = 75^\circ$, Sect.~\ref{sect:resfid}). The orange shaded region is the observational constraint on $\psi$. An aggregated histogram of the final spin--orbit angles is also shown on the far right.}
\label{fig:gj436_psif}
\end{figure*}

\end{appendix}

\end{document}